\documentclass[a4paper,twocolumn,11pt]{quantumarticle}
\pdfoutput=1
\usepackage[utf8]{inputenc}
\usepackage[english]{babel}
\usepackage[T1]{fontenc}
\usepackage{hyperref}
\usepackage{cite}

\usepackage{amsmath, amssymb}
\usepackage{graphicx}
\graphicspath{{./graphics/}}
\usepackage{dcolumn}
\usepackage{bm}
\usepackage{qcircuit}
\usepackage{mathtools}
\usepackage{physics}
\usepackage{dsfont}
\usepackage[dvipsnames]{xcolor}
\usepackage{hyperref}
\usepackage{faktor}
\usepackage{subfigure}

\usepackage{tikz}
\usepackage{lipsum}

\DeclareMathOperator{\spn}{span}
\newcommand{\zz}{\mathbb{Z}}

\newcommand\numberthis{\addtocounter{equation}{1}\tag{\theequation}} 

\usepackage[numbers,sort&compress]{natbib}

\begin{document}

\title{Non-Pauli topological stabilizer codes from twisted quantum doubles}
\author{Julio Carlos Magdalena de la Fuente}
\email{jm@juliomagdalena.de}
\affiliation{JARA Institute for Quantum Information, RWTH Aachen University, Aachen, Germany}
\affiliation{Dahlem Center for Complex Quantum Systems, Freie Universit{\"a}t Berlin, 14195 Berlin, Germany}%
\author{Nicolas Tarantino}%
\affiliation{Dahlem Center for Complex Quantum Systems, Freie Universit{\"a}t Berlin, 14195 Berlin, Germany}%
\author{Jens Eisert}
\affiliation{Dahlem Center for Complex Quantum Systems, Freie Universit{\"a}t Berlin, 14195 Berlin, Germany}%

\maketitle

\begin{abstract}
It has long been known that long-ranged entangled topological phases can be exploited to protect quantum information against unwanted local errors. Indeed, conditions for intrinsic topological order are reminiscent of criteria for faithful quantum error correction. At the same time, the promise of using general topological orders for practical error correction remains largely unfulfilled to date. In this work, we significantly contribute to establishing such a connection by showing that Abelian twisted quantum double models can be used for quantum error correction. By exploiting the group cohomological data sitting at the heart of these lattice models, we transmute the terms of these Hamiltonians into full-rank, pairwise commuting operators, defining commuting stabilizers. The resulting codes are defined by commuting non-Pauli stabilizers, with local systems that can either be qubits or higher dimensional quantum systems. Thus, this work establishes a new connection between condensed matter physics and quantum information theory, and constructs tools to systematically devise new topological quantum error correcting codes beyond toric or surface code models.
\end{abstract}

\section{Introduction}
Any architecture proposed for information storage must be equipped with an error correction strategy to avoid the corruption of the data encoded, whether the information is classical or quantum in nature \cite{QEC,SteaneCode,CalderbankShor}. Since the no-cloning
theorem \cite{nielsen2002quantum} prevents qubits from being copied,
quantum error correction cannot rely on simply copying the necessary information at any point. Thankfully, the fact that errors are usually local, i.e., they affect a small number of qubits, has lead to fruitful alternative strategies.
By distributing the relevant data over a whole system, it is possible to detect the errors without ever needing to copy the original state.

Building from this insight, stabilizer codes \cite{gottesman1997stabilizer,RevModPhys.87.307} have taken a particularly prominent role in the search for encoding strategies for
scalable and fault-tolerant quantum computing. In stabilizer codes, the subspace
in which the quantum information is stored is the joint eigenspace of pairwise commuting operators, called \textit{stabilizers}. Among these are a class of  codes -- so called \textit{topological} codes -- where error detection can be performed with the measurement of \textit{local} stabilizers. These measurement outcomes, repackaged into \textit{syndromes}, determine the errors that have occurred. By construction of such codes, the measurement does not destroy the stored quantum information and makes it possible to restore it with a suitable error correction scheme \cite{gottesman1997stabilizer,TopologicalQuantumMemory,RevModPhys.87.307}.
The \emph{toric code} \cite{kitaev2003fault},  its associated planar embedding known as the \emph{surface code} \cite{PhysRevA.86.062318}, and \emph{color codes} \cite{PhysRevLett.97.180501}, are by far the most studied codes, and have emerged as the gold standard of this class of error-correction protocols. Their simple construction -- with stabilisers built out of Pauli words -- means that they collectively provide a wide range of easily understood schemes. That said, there are strong reasons to seek for new codes beyond these Pauli stabilizer models.
While the lack of a universal
and fault-tolerant gate set -- by virtue of the
\emph{Eastin-Knill theorem} \cite{nielsen2002quantum} --
and a lack of self-correctability \cite{BravyiTerhalNoGo} will
be common to any stabilizer approach in two spatial dimensions, several
techniques have already been identified to circumvent these limitations, including \emph{magic state distillation} \cite{Litinski} and
just-in-time decoders \cite{Brown,Bombin}.
Moreover, codes 
built out of $d$-level systems have
been found to have
superior error correction capabilities compared to qubit-based
codes, with an increasing performance with increasing $d$ \cite{PhysRevLett.113.230501} or enhanced bit flip stability \cite{andrist2015error}.
%
%
%
Other generalisations involving non-commuting stabiliser sets
\cite{XS} have demonstrated the ability to produce gate sets which, while not universal, have enhanced computation power.
Taken together, these findings strongly motivate the quest for
new topological quantum error correction codes with stabilizers outside the Pauli group.


In light of this search, we present a wealth of new topological
codes. To do so, we have taken inspiration from the closely related field of topological phases of matter. The conditions for quantum error correction, the
Knill-Laflamme criteria \cite{knill1997theory}, are highly reminiscent of conditions
for the topological order in quantum many-body theory. However, this connection is rarely made explicit beyond the toric code, which can be seen
as defining a gapped Hamiltonian with 4 anyon types and a topological ground state
degeneracy.
While it is true that all topological error correcting
codes can ultimately be understood as defining a system containing anyonic excitations and therefore being in a topological phase, all well-studied instances of this are equivalent to multiple copies of the toric code phase \cite{bombin2012universal}.
What is sorely lacking in this picture is a way of reverse-engineering topological quantum
error correcting codes from the wealth of topological phases of matter.
This seems a remarkable omission in the light of the powerful and highly developed classification of such phases from the perspective of condensed matter and mathematical physics \cite{levin2005string, hu2013twisted, chen2013symmetry}. This omission is also significant given the fact that, from a technological perspective \cite{Roadmap}, the identification of new topological codes seems imperative.

In this work, we use a large class of topological orders hosting Abelian anyons to construct new topological error correcting codes.
In particular, we modify existing lattice models for topological orders -- twisted quantum double models -- so that they give rise to Non-Pauli stabilizers. In their original form, the local terms of these Hamiltonians do not commute in a particular excited subspace of the Hilbert space, which makes them -- on first glance -- unsuitable for stabilizer error correction. Practically speaking, commutativity is a highly desirable property in the context of quantum error correction, in that it allows for error correction schemes based on \emph{independent} local measurements of such stabilizers without perturbing the stored quantum information. We restore commutativity by first deriving the quantities that obstruct this property from the \emph{group cohomology} data of twisted gauge theories. In most cases -- namely, for Abelian twisted quantum doubles -- these obstructions can be lifted completely by carefully modifying the offending terms in the Hamiltonian, yielding a true stabilizer code, consisting of commuting \emph{non-Pauli} operators.
A first step in this direction was taken in Ref.~\cite{dauphinais2019quantum}, where the \emph{double semion string-net model} \cite{levin2005string} was modified with a local phase factor to overcome the same commutativity problem. However, this
approach lacks a systematic and quantitative understanding of the
failure of commutativity, and as such it cannot be generalized to other lattice models for more exotic topological orders. Our results go a significant step further, providing a robust framework for deriving quantum error correcting codes from not only a Hamiltonian in the double semion phase, but from a huge family of Abelian phases as well. The main limitation of our scheme is that it is not applicable to TQD models with Non-Abelian anyons.

This work is structured as follows: In Section\,\ref{sec:introTQD} we give a comprehensive introduction to twisted quantum double models for general (Abelian) groups. We have kept the mathematical details to a minimum while still presenting our results in a self-contained manner. Our construction is done explicitly for a $\zz_2$ and a $\zz_2\times\zz_2$ model and then summarized for the general $\zz_N$ and $\zz_N^2$ cases in Sec.\,\ref{sec:fullycommutingmodels}, with full details in Apps.\,\ref{app:obstructions}-\ref{app:ZNZN}. Moreover, we give a brief overview on the properties of the newly constructed codes and how they relate to known schemes for topological quantum error correction. In Sec.\,\ref{sec:conclusion}, we conclude our work and give an outlook on future directions and potential use cases of the codes and discuss potential applications in topological quantum information processing.

\section{Introduction into twisted quantum double models}\label{sec:introTQD}

\textit{Twisted quantum double (TQD)} models are lattice models for topological order in 2+1 dimensions which can be viewed as a generalization of the quantum double model \cite{kitaev2003fault} introduced by Kitaev. They can be obtained by promoting the global symmetry of a \emph{symmetry protected topological (SPT) phase} \cite{chen2013symmetry, levin2012braiding, chen2017symmetry} to a \emph{local gauge symmetry} via minimal coupling to the original ``spins'' of the SPT. We will restrict the discussion of the model only to the aspects necessary for our construction. An interested reader is referred to Ref.~\cite{hu2013twisted} for more comprehensive perspective.

\begin{figure}[t]
  \centering
  \includegraphics[width=0.35\textwidth]{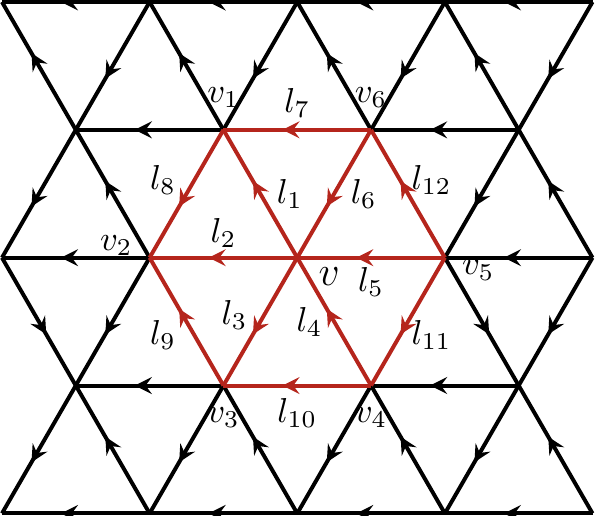}
  \caption{Local patch of a translation-invariant lattice on which we define our model. It is a oriented triangulation of a compact surface. Around each vertex, the edges directly adjacent to it are labelled from $l_1$ to $l_6$ and the other 6 edges that share a triangle with the vertex are labelled by $l_7$ to $l_{12}$. Together, these edges constitute the \emph{neighborhood} of the vertex and are marked in red above.}
  \label{fig:lattice}
\end{figure}

\subsection{The Hamiltonian}
We define our model on a translation-invariant, oriented triangulation of a general compact surface shown in Fig.\,\ref{fig:lattice}.
We label the edges in the neighborhood of a vertex $v$ from $l_1$ to $l_{12}$. Each edge $l_i$ carries a degree of freedom (gauge field) whose local Hilbert space $\mathcal{H}_l$ is spanned by states labeled by elements of a finite group $G$,
\begin{align}
  \mathcal{H}_l = \spn_{\mathbb{C}}\{\ket{g},\;g\in G\}
  \label{eq:hilbdef}
\end{align}
with $\braket{g}{h}=\delta_{g,h}$. Its local dimension is $|G|$, so a group with $|G|=2$ will be a qubit model , $|G|=3$ a qutrit model and so on. The total Hilbert space is then simply given by $\mathcal{H} = \bigotimes_{\text{edges }l}\mathcal{H}_l$. While $G$ can be chosen to be any finite group, we will only be treating Abelian cases in this work.
Additionally, the TQD model takes a cocycle function $\omega: G^2 \to U(1)$ as an input.\footnote{To be precise, the cocycle $\omega$ has to be normalized, i.e. $\omega(0_G,a,b) = \omega(a,0_G,b) = \omega(a,b,0_G)=1\;\forall a,b\in G$.} It will define the action of the \textit{vertex operators} in the Hamiltonian.

The full TQD Hamiltonian is given by
\begin{align}
  H_{TQD} = -\sum_{\text{plaquettes }p}B_p - \sum_{\text{vertices }v}A_v.
\end{align}
The first sum runs over all (triangular) plaquettes and the \textit{plaquette operator} acting on a triangular face is defined by
\begin{align}\label{eq:plaquette_projector}
  \begin{split}
    B_p &\Bigg| \begin{minipage}[c]{1.3cm}
      \includegraphics[width=\textwidth]{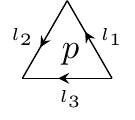}
    \end{minipage}  \Bigg\rangle\\
    &= \delta_{(-l_3)\oplus l_2\oplus l_1} \Bigg| \begin{minipage}[c]{1.35cm}
      \includegraphics{triangle_labeled}
    \end{minipage} \Bigg\rangle\\
    &= \frac{1}{\abs{G}} \sum_{i\in C} \underbrace{\chi_i(-l_3 \oplus l_2\oplus l_1)}_{\eqqcolon B_p^i}\Bigg| \begin{minipage}[c]{1.3cm}
      \includegraphics[width=\textwidth]{triangle_labeled}
    \end{minipage}  \Bigg\rangle,
  \end{split}
\end{align}
where $\oplus$ denotes the group operation (modular addition for cyclic groups) and $-l$ the inverse element of $l$ in $G$. The projector $\delta_g = 1$ for $g=0_G$ (identity element in $G$) and $\delta_g = 0$ otherwise. The plaqeutte terms ensure that the ground space is flux free, i.e., $B_p=1\;\forall p$. In the second step, we decomposed the projector into a sum of group characters $\chi_i$ over conjugacy classes $C$, which can be done for any finite group.
While this decomposition is usually a curiosity, this is useful when $G$ is an Abelian group. In this case, each conjugacy class contains only a single element, and so the characters are in one-to-one correspondence with the group elements. For cyclic groups, where $g \in \{0,\dots,N-1\}$, this means that these character operators can be thought of as the products \emph{phase operators} $Z_l$,
\begin{subequations}\label{eq:Bp_decomposition}
  \begin{align}
    B_p^g =& \prod_{i\sim p}\sum_{n=0}^{N-1}\left(e^{\frac{2\pi i}{N}n}\right)^{s(p,l)g}\ketbra{n}\\
    =& \prod_{l\sim p}Z_l^{s(p,l) g},
  \end{align}
\end{subequations}
with $s(p,i) = +1(-1)$ if edge $l$ is oriented in counterclockwise(clockwise) direction around plaquette $p$. For $G=\zz_2$ for example, $Z_i$ is the Pauli $Z$ operator acting on edge $i$ and one recovers the toric code plaquette operator. Arbitrary finite abelian groups can always be decomposed into cyclic factors, and thus their character operators can be factored into products of phase operators as well.

The second sum in the Hamiltonian runs over all vertices and the summand $A_v$, acting on the neighborhood of a vertex $v$, is defined by a similar decomposition as $B_p$, i.e.,
\begin{align}
  A_v = \frac{1}{\abs{G}}\sum_{g\in G}A_v^g,
\end{align}
where $\abs{G}$ is the order of the group $G$, and $A_v^g$ the \emph{vertex operator} corresponding to the group element $g$ defined by its action on a basis element,
\begin{widetext}
\begin{align}
  A_v^g \Bigg| \begin{minipage}[c]{2.2cm}
  \includegraphics[width=\textwidth]{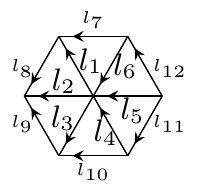}
\end{minipage}\dots
\Bigg\rangle =& \frac{\omega(l_9,l_3',g) \omega(l_3', g,l_4) \omega(g,l_4,l_{11})}{\omega(l_8,l_1',g) \omega(l_1',g,l_6) \omega(g,l_6,l_{12})} \Bigg| \begin{minipage}[c]{2.2cm}
 \includegraphics[width=\textwidth]{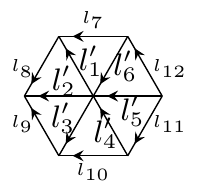}
\end{minipage}\dots \Bigg\rangle,\label{eq:vertex_action}
\end{align}
\end{widetext}
where the label of the central vertex $v$ is left out for readability. The vertex operator can be decomposed into two consecutive actions. First, it changes the values of the edges adjacent to the vertex depending on their orientation.
On our lattice,
\begin{subequations}\label{eq:vertex_action_edges}
\begin{align}
  &l_i \stackrel{A_v^g}{\longmapsto} l_i^\prime = l_i\oplus (-g) & \text{for } i=1,2,3,\\
  &l_j \stackrel{A_v^g}{\longmapsto} l_j^\prime=g\oplus l_j & \text{for } j=4,5,6,
\end{align}
\end{subequations}
where $\oplus$ denotes the group operation which we choose to be modular addition for cyclic groups.
Second, it scales the wavefunction by a phase factor given by the product of $6$ \textit{cocycles} $\omega: G^3 \to U(1)$, one per triangle adjacent to $v$. The order of the arguments in the cocycles and whether they appear in the numerator or denominator of this pre-factor is determined by the orientation structure of the lattice. For a detailed explanation of constructing the pre-factor for a general lattice, see Ref.\,\cite{hu2013twisted}. The cocycles encode the \emph{topological data} of the theory modelled by $H_{TQD}$. Their defining property is the so-called \textit{cocycle condition}
\begin{align}
\begin{split}
  \frac{\omega(g_1,g_2,g_3)\omega(g_0,g_1\oplus g_2,g_3)\omega(g_0,g_1,g_2)}{\omega(g_0\oplus g_1,g_2,g_3)\omega(g_0,g_1,g_2\oplus g_3)} = 1 \\[6pt]
  \forall g_0,g_1,g_2,g_3\in G. \label{eq:cocycle_condition}
\end{split}
\end{align}
Obviously, $\omega(a,b,c)\equiv1$ is always a solution and is called trivial. If we use this trivial solution in Eq.\,\eqref{eq:vertex_action} to define $H_{TQD}$, we obtain the
\emph{quantum double Hamiltonian} from \cite{kitaev2006anyons}. Since -- in general -- there are non-trivial solutions to this equation as well, the TQD model covers a much broader class of Hamiltonians than the pure quantum double Hamiltonians. In principle, one can choose any function satisfying condition \eqref{eq:cocycle_condition}, insert it into Eq.\,\eqref{eq:vertex_action} and obtain a consistent topologically ordered Hamiltonian. However, not all solutions yield distinct orders but they are classified in equivalence classes. A close investigation of the cocycle condition reveals that if we have one solution $\omega$, we can always obtain another solution
\begin{align}
  \Tilde{\omega}(g_1,g_2,g_3) = \omega(g_1,g_2,g_3) \frac{\beta(g_2,g_3)\beta(g_1,g_2\oplus g_3)}{\beta(g_1\oplus g_2,g_3) \beta(g_1,g_2)},\label{eq:coboundary_gauge}
\end{align}
where $\beta: G^2\to U(1)$ is an arbitrary function mapping two group elements to a phase factor. If we have two TQD Hamiltonians defined by two cocycles $\omega_1$ and $\omega_2$ in Eq.\,\eqref{eq:vertex_action} so that they are in different topological orders, we know and there exists no $\beta$ to map $\omega_1$ onto $\omega_2$ by Eq.\,\eqref{eq:coboundary_gauge}.
Hence, inequivalent Hamiltonians $H_{TQD}$ (in the sense of topological order)
are classified by distinct equivalence classes of functions $\omega$, which define elements of the third group cohomology of $G$ over $U(1)$
\begin{align}
  [\omega]\in H^3(G,U(1))= \faktor{\{\omega\text{ satisfying }\eqref{eq:cocycle_condition}\}}{\sim}\label{eq:formalclassification}
\end{align}
with $\omega \sim \Tilde{\omega}$ iff $\exists\beta:G^2\to U(1)$ such that they are related by Eq.\,\eqref{eq:coboundary_gauge}. In the next section, we will see examples of such functions for simple groups such as $\zz_2$ and $\zz_2\times \zz_2$.
For an introduction into group cohomology, see App.\,\ref{app:group_cohomology}.

\subsection{Topological data}
The Hamiltonian we have constructed on a triangulation of a compact surface from a group $G$ together with a cocycle $\omega$ is indeed topologically ordered. It has anyonic excitations and a robust ground state degeneracy (GSD). For models with Abelian anyons only, $\text{GSD} = \abs{G}^{2g}$ where $g$ is the genus of the surface on which it is defined. The topological quantum numbers (topological spin and S-matrix) of the excitations are uniquely defined by the input group $G$ and the cocycle one chooses to define $A_v$. Moreover, they are gauge invariant in the sense that one can choose any cocycle in the same equivalence class to define $A_v$ and still obtain the same topological data. This corresponds to a transformation of the cocycles like in Eq.\,\eqref{eq:coboundary_gauge}. For the derivation and explicit expressions of those quantities in terms of the input data see Ref.~\cite{hu2013twisted}.

\subsection{Ground space and failure of commutativity}
The ground space of $H_{TQD}$ can be found exactly and is defined implicitly by the conditions
\begin{align}
  \begin{split}
    B_p\ket{\psi} = A_v\ket{\psi} = \ket{\psi}\quad \forall &\text{ plaquettes }p\\
    &\text{ and vertices } v,
  \end{split}
\end{align}
so that it is the simultaneous eigenspace of the plaquette and vertex operators. As stated before, this subspace has dimension larger than 1 on a surface with non-trivial topology and therefore we hope to use that space as a code space of an error correction stabilizer code. Unfortunately, the Hamiltonian is not exactly solvable on the whole Hilbert space, i.e. one cannot simultaneously diagonalize all vertex and plaquette operators. In particular, the vertex operators fail to commute in the presence of certain fluxes ($B_p=0$).\footnote{For a precise statement on how commutativity fails for general Abelian groups, we refer to App.\,\ref{app:obstructions}.}
In the TQD model for $G=\zz_2$ and the non-trivial cocycle \cite{hu2013twisted}
\begin{align}\label{eq:Z2_cocycle}
    \omega_1(1,1,1) = -1, \qq{} \omega(a,b,c) = 1 \quad\text{else,}
\end{align}
a version of the double semion phase, the vertex operators do not commute when acting on the following configuration:
\begin{align*}
    A_{2}^1 A_v^1\ket{\begin{minipage}[c]{1.5cm}
     \includegraphics[width=\textwidth]{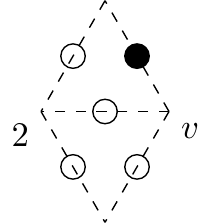}
    \end{minipage}}
    = \ket{\begin{minipage}[c]{1.5cm}
      \includegraphics[width=\textwidth]{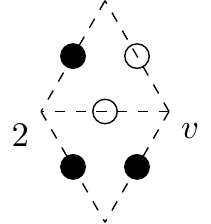}
    \end{minipage}}\\
    A_v^1 A_{2}^1\ket{\begin{minipage}[c]{1.5cm}
      \includegraphics[width=\textwidth]{flux_config1}
    \end{minipage}}
    = -\ket{\begin{minipage}[c]{1.5cm}
      \includegraphics[width=\textwidth]{flux_config3}
    \end{minipage}}.
  \end{align*}
In the case of $G=\zz_2$, the local Hilbert space corresponds to the one of a qubit which we represented by a circle in the state vectors above. We have labelled the state vectors of the qubits with circles, $\ket{0}=\circ$ and $\ket{1}=\bullet$. The rest of the lattice is not explicitly shown, but we assume that all other qubits are in $\ket{0}$. The vertex operators will be defined in Sec.\,\ref{sec:Z2} where we discuss the double semion phase in detail.

From an \emph{error correction perspective}, there are many (single qudit) errors that create such fluxes and for the vertex operators to be proper stabilizers they also have to commute in that sector of the Hilbert space. In principle, one can make the Hamiltonian exactly solvable by multiplying it with a projector on the flux-free subspace, $P_{B=1} = \prod_p B_p$. However, we will loose information about the excited sector in doing so in the sense that it makes it impossible to identify an excitation uniquely by measuring those local operators. Luckily, it turns out that a slight modification of the vertex operators that does not change the topological order resolves the obstruction of commutativity entirely.

\section{Construction of fully commuting models}\label{sec:fullycommutingmodels}
In order to successfully overcome the obstacle to commutativity, the vertex operators have to be modified.
This modification should not alter the \emph{topological phase} of the Hamiltonian.
This means that the vertex operators must be altered in such a way so that the ground space is left unchanged and that the spectral gap is preserved. At the same time, the modification should be \emph{minimal}.
The latter means that the modification should be local -- in the sense of not increasing the support at all -- and leave the large sections of the Hilbert space where the Hamiltonian \emph{is} solvable undisturbed. In particular, only for certain flux configurations do the operators need to be minimally altered.

It constitutes the main result of this work that, for
large classes of
twisted quantum doubles, the above desiderata can be achieved with a modification of the form,
\begin{align}\label{eq:general_phasemodification}
    \Tilde{A}_v^g = D_v^gA_v^g,
\end{align}
where $D_v^g$ is a full-rank operator that is diagonal in the edge basis, with entries being roots of unity, and is equal to $\mathds{1}$ on the ground space.

By imposing that the modified operators $\{\Tilde{A}_v^g\}$ commute, we obtain consistency equations for the modification phases $\{\eta_v^g\}$. The derivation of these consistency equations is \emph{only} possible with the TQD model and exploits the machinery of group cohomology in its construction.
We will illustrate the procedure of solving these equations by means of two simple examples where the input group is $\zz_2$ and $\zz_2\times \zz_2$. In the second case, we investigate a topological order that is entirely new to the context of quantum error correction, but is still in principle realizable with a qubit architecture. We apply the same procedure to more exotic models derived from the groups $\zz_N$ and $\zz_N^2$. With those completed, we have resolved the commutativity issue for every Abelian topological order that can be obtained from a twisted quantum double model. These results are described in the last part of this section. For the general formalism and the calculation see App.\,\ref{app:obstructions}, \ref{app:ZN}, and \ref{app:ZNZN}.

The newly obtained operators $\Tilde{A}_v^g$ together with the plaquette phase operators $B_p^g$ (see Eq.\,\eqref{eq:plaquette_projector}) generate the Non-Pauli stabilizer group
\begin{equation}
    \mathcal{S}_{TQD} = \langle \{B_p^g\}, \{\Tilde{A}_v^g\} \rangle,
\end{equation}
which defines the corresponding code. Although stabilizer groups are normally defined as a subgroup of the Pauli group, this restriction is not necessary for the general concepts of stabilizer error correction. These require \emph{only} that there exist an invariant subspace under the action of those generalized stabilizers, namely the common +1 eigenspace of all $B_p^g$ and $\Tilde{A}_v^g$.

In the following, we will illustrate the construction for an arbitrary finite group $G$ and a cocycle $\omega$ for two simple examples that already go beyond the toric code. We will see that we arrive at a wealth of new topological quantum error correcting codes produced directly from twisted quantum double models for topological order.

\subsection{$\zz_2$ -- double semion code}\label{sec:Z2}
We first investigate the non-trivial $\zz_2$ model that is in the same phase as the double semion string-net model
\cite{levin2005string}. We represent $\zz_2$ as the set $\{0,1\}$ together with the group operation being addition modulo two. With that, $\mathcal{H}_l = \spn_\mathbb{C}\{\ket{0},\ket{1}\} = \mathbb{C}^2$, so it is a model of interacting qubits.
$\zz_2$ has two inequivalent cocycle classes, one trivial class $[\omega_0\equiv 1]$ and one non-trivial class $[\omega_1]$. The TQD model with the trivial cocycle would yield a Hamiltonian in the toric code phase. The canonical representative of the non-trivial class is given by Eq.\,\eqref{eq:Z2_cocycle}.
Inserting this cocycle into Eq.\,\eqref{eq:vertex_action} yields the Hamiltonian
\begin{widetext}
\begin{subequations}
\begin{align}
    H_{\zz_2} =& -\sum_{\text{plaquettes }p} \frac{1}{2}\Big(\mathds{1}
    + \prod_{l\sim p} Z_l \Big)
    - \sum_{\text{vertices }v} \frac{1}{2}\Big(\mathds{1} + A_v^1\Big)\\
    &\qq{with}
    A_v^1 = (-1)^{P_9^-P_3^- + P_3^-P_4^+ + P_4^+P_{11}^- +P_8^-P_1^- + P_1^-P_6^+ + P_6^+P_{12}^-} \prod_{l=1}^6X_l,
\end{align}
\end{subequations}
\end{widetext}
where $Z_l$ and $X_l$ are the Pauli $Z$ and $X$ matrices on (the qubit sitting on) edge $l$ and $P_l^{\pm} = \frac{1}{2}(1\pm Z_l)$ is the projector onto the space where edge $l$ caries the value $0$ or $1$, respectively. Each term of the form $(-1)^{P^-_i P^-_j}$ corresponds to a $CZ$ gate between qubit $i$ and $j$. These entangling gates are essential to build up long range entanglement that is inequivalent to the one in the toric code. The above Hamiltonian is in the so-called \emph{double semion phase}.

By construction, $A_v^1$ always flips an even number of qubits adjacent to a plaquette, and thus $\comm{A_v^1}{B_p} = 0\;\forall v,p$.\footnote{In fact, $A_v^g$ is flux preserving in any twisted quantum double model due to the orientation structure of the edges.} Hence, the only obstruction for the operators $B_p$ and $A_v^1$ to form a commuting set of operators that we can use for stabilizer error correction comes from the vertex operators $A_v^1$.

\subsubsection{Obstruction in the original model}
For the operators $A_v^1$ to generate a stabilizer group they have to represent the group action of $\zz_2$ on site. In particular, any representative of an element in $\zz_2$ should square to the identity. Unfortunately, it turns out that
\begin{align}
\begin{split}
    \left(A_v^1\right)^2 =& (-1)^{P_1^- + P_3^- + P_4^- + P_6^- + P_8^- + P_9^- + P_{11}^- + P_{12}^-}\mathds{1}\\
    \neq& \mathds{1},\label{eq:Z2_onsiteobstruction1}
\end{split}
\end{align}
where we have used $X_l^2 = \mathds{1}$, and the decomposition of the identity, $\mathds{1} = P_l^+ + P_l^-$ for any edge $l$. In fact, $\left(A_v^1\right)^2 = -\mathds{1}$ exactly when an odd number of the edges $\{1,3,4,6,8,9,11,12\}$ is in the state $\ket{1}$, which coincides with
\begin{align}
    \left(A_v^1\right)^2 = (-1)^{B_{3,9,8,1} + B_{6,12,11,4}}\mathds{1}, \label{eq:Z2_onsiteobstruction2}
\end{align}
where $B_{i,j,k,l} = \frac{1}{2}(1-Z_iZ_jZ_kZ_l)$ measures the flux through the region enclosed by the edges $\{i,j,k,l\}$.
This shows that the operators fail to represent the group action on the part of the Hilbert space where $B_{3,9,8,1} + B_{6,12,11,4} = 1\mod 2$. In particular, on the ground space (in which no flux is present) the group action is implemented correctly.

This is not the only obstruction in the original TQD model. To form stabilizers, the vertex operators $A_v^g$ must commute pairwise. Due to the translational invariance of the model, we need only calculate the commutation relation between $A_v^1$ with the three operators $\{A_i^1,\;i=1,2,3\}$ acting on the three vertices $\{1,2,3\}$ connected to $v$ by the edges $\{l_1,l_2,l_3\}$ (see Fig.\,\ref{fig:lattice} for the labelling) to confirm this. It turns out that
\begin{subequations}\label{eq:Z2_commutationobstructions}
\begin{align}
    A_{v_1}^1A_v^1 =& A_v^1A_{v_1}^1,\\
    A_{v_3}^1A_v^1 =& A_v^1A_{v_3}^1,\\
    \begin{split}
        A_{v_2}^1A_v^1 =& (-1)^{P_1^-+P_3^-+P_8^-+P_9^-}A_v^1A_{v_2}^1\\
        =& (-1)^{B_{3,9,8,1}}A_v^1A_{v_2}^1,
    \end{split}
\end{align}
\end{subequations}
using the same relations used to produce Eq.\,\eqref{eq:Z2_onsiteobstruction1} and \eqref{eq:Z2_onsiteobstruction2}. Again, we find that they commute in the zero-flux sector of the Hilbert space. Interestingly, vertex operators on neighboring vertices only fail to commute in the last case, when they are connected by a horizontal edge (labeled by $l_2$ in Fig.\,\ref{fig:lattice}) which is neighboring a nontrivial flux $B_{3,9,8,1} = 1 \mod 2$. This particular ``locality" of the commutativity obstruction is a consequence of our chosen edge orientation\footnote{Choosing a different edge orientation in the first place would only shift the commutativity obstruction to different edges, not remove it.} which determines which arguments enter in the cocycles in Eq.\,\eqref{eq:vertex_action}.

\subsubsection{Modifying vertex operators by local phase}
We have found that the vertex operators in the original TQD model fail to be commuting stabilizers on the whole Hilbert space because, on one hand, they do not implement the group action on site, i.e. $(A_v^1)^2\neq \mathds{1}$, and, on the other, fail to commute $A_{v_1}^1A_{2}^1\neq A_{2}^1A_{v_1}^1$. However, we were able to quantify the obstructions and found that they take a very particular form and only depend on fluxes. Because of this, there are no obstructions in the ground space. To remove the obstructions on the whole Hilbert space, we want to modify the vertex operators by a local diagonal unitary as described in the beginning of Sec.\,\ref{sec:fullycommutingmodels},
\begin{align}
    \Tilde{A}_v^1 = D_v A_v^1
\end{align}
so that $\Tilde{A}_v^1$ are stabilizers. The modified operators should therefore satisfy
\begin{subequations}
\begin{align}
    \left(\Tilde{A}_v^1\right)^2 =& \mathds{1},\label{eq:Z2_condition1}\\
    \Tilde{A}_{v_1}^1\Tilde{A}_{v_2}^1 =& \Tilde{A}_{v_2}^1\Tilde{A}_{v_1}^1, \label{eq:Z2_condition2}
\end{align}
\end{subequations}
for all vertices $v,v_1,v_2$. Imposing Eq.\,\eqref{eq:Z2_condition1} and using Eq.\,\eqref{eq:Z2_onsiteobstruction2} yields the condition on the modification phase
\begin{align}
    1 = D_v A_v^1D_v \left(A_v^1\right)^{-1} (-1)^{B_{3,9,8,1} + B_{6,12,11,4}}.
\end{align}
In fact, this can be solved by the Ansatz
\begin{align}
    D_v = i^{B_{3,9,8,1} + B_{6,12,11,4}} \bar{D}_v
\end{align}
with $\bar{D}_v A_v^1\bar{D}_v \left(A_v^1\right)^{-1} = 1$ since $A_v^1$ leaves fluxes invariant. Now, inserting this family of solutions into Eq.\,\eqref{eq:Z2_condition2} and using Eqs.\,\eqref{eq:Z2_commutationobstructions} yields a second condition on $\bar{D}_v$,
\begin{align}
\begin{split}
    1 =& (-1)^{B_{3,9,8,1}}\bar{D}_2 A_v^1(\bar{D}_2)^{-1}\left(A_v^1\right)^{-1}\\
    \times& (\bar{D}_v)^{-1} A_2^1\bar{D}_v \left(A_2^1\right)^{-1},
\end{split}
\end{align}
which is solved by the one-parameter family
\begin{align}
\begin{split}
    \bar{D}_v(p) =& e^{\pi i [pB_{3,9,8,1}L_2 + (p-1)B_{6,12,11,4}L_5]}\\
    \times& e^{-\frac{\pi i}{2}[(p+2)B_{3,9,8,1} + (p+1)B_{6,12,11,4}]},
\end{split}
\end{align}
with $p\in\mathbb{R}$. $L_i = \frac{1}{2}(1-Z_i)$ is the operator that measures the value of the edge $l_i$ in $\zz_2$. Note that $\bar{D}$ is a periodic function in $p$ with periodicity $4$, i.e. $\bar{D}(p+4)=\bar{D}(p)$. Also, the second factor containing $i$ only depends on fluxes and ensures that $\bar{D}_v A_v^1\bar{D}_v \left(A_v^1\right)^{-1} = 1$ and does not affect commutativity.

\begin{figure}
    \centering
    \includegraphics[width=0.3\textwidth]{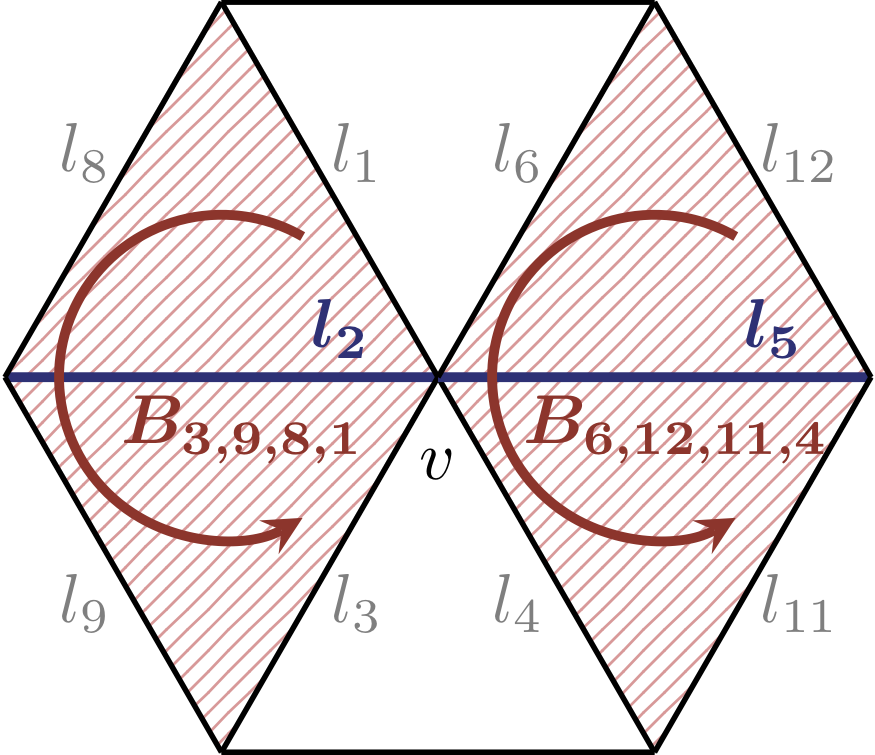}
    \caption{Geometric illustration of the support of $D_v^1(p)$ in the neighborhood of vertex $v$. For any $\zz_N$ model (see App.\,\ref{app:ZN}), the obstructions (and therefore also the modification phases) only depend on the horizontal edges, labelled by $l_2$ and $l_5$ and the two fluxes through the regions around them, $B_{3,9,8,1}$ and $B_{6,12,11,4}$, depicted by a dashed area in the figure above. The edges $l_1,l_8,\dots$ that are written in gray only enter implicitly via the fluxes that are composed of them. Note the orientation structure can be neglected in the $\zz_2$ case.}
    \label{fig:Z2_supportstructure}
\end{figure}
Combined, we obtain the modification phase
\begin{align}
\begin{split}
    D_v(p) =& e^{-\frac{\pi i}{2}[(p+1)B_{3,9,8,1} + pB_{6,12,11,4})]}\\
    \times& e^{\pi i[pB_{3,9,8,1}L_2 + (p-1)B_{6,12,11,4}L_5]},
\end{split}\label{eq:Z2_generalcorrection}
\end{align}
where the first factor ensures that the group property is fulfilled on site and the second factor fixes the commutativity on the whole Hilbert space. $D_v(p+4)=D_v(p)$, $p\in[0,4)$ parametrizes all the distinct modification phases in this family of solutions. The geometric support structure for general $p$ is depicted in Fig.\,\ref{fig:Z2_supportstructure}. Note that the parameter $p$ sets the dependence on $l_2$ and $l_5$ in the second factor. The freedom to choose $p\in[0,4)$ may be useful in an actual error correction scheme since different modification phases yield different stabilizers that in turn could have different properties in the decoding process.
For $p=1$, for example, we obtain
\begin{align}\label{eq:Z2_correctionphasep1}
    D_v(1) =& (-i)^{B_{6,12,11,4}}(-1)^{B_{3,9,8,1}}(-1)^{B_{3,9,8,1}L_2},
\end{align}
so that it does not depend on $l_5$.

Explicitly quantifying the obstructions of the group property (Eq.\,\eqref{eq:Z2_onsiteobstruction2}) and the commutativity of the operators acting on neighboring vertices (Eqs.\,\eqref{eq:Z2_commutationobstructions} in the original TQD model enabled us to remove them with a local phase modification such that the modified operators $\Tilde{A}_v^1 = D_vA_v^1$ faithfully represent the group $\zz_2$ on the whole Hilbert space. The constructed operators are outside the Pauli group and can be used as stabilizers in the context of quantum error correction. Moreover, the modification does not change the action of $A_v^1$ on the ground space, and thus the modified Hamiltonian it still is in the double semion phase.

\subsection{$\zz_2\times \zz_2$ -- twisted color codes}\label{sec:Z2Z2}
In the previous section, we have constructed a set of stabilizers defined on a lattice of qubits such that its code space corresponds to a double semion ground space. However, the double semion phase is not the only twisted gauge theory one can implement with a qubit architecture. By taking $G=\zz_2\times\zz_2 = \{g=(g_1,g_2);\;g_1,g_2\in\zz_2\}$ as an input group for the TQD model, the local Hilbert space becomes $\mathcal{H}_l=\spn_\mathbb{C}\{\ket{0,0}, \ket{0,1}, \ket{1,0}, \ket{1,1}\}\simeq \mathbb{C}^2\otimes\mathbb{C}^2$ which can be realized using two qubits per edge. As was shown in Ref.\,\cite{kubica2015unfolding}, an untwisted $\zz_2\times\zz_2$ quantum double model -- otherwise known as two copies of the toric code -- is equivalent to the color code. In this section we investigate the twisted versions thereof, which we call \textit{twisted color codes}.

The possible topological orders of a TQD model with $G=\zz_2\times\zz_2$ are classified by $H^3(\zz_2\times\zz_2,U(1))=\zz_2\times\zz_2\times\zz_2$, and thus there are 8 different cocycle classes we can choose as input, labelled by $(s_1,s_2,s_3)\in\zz_2^3$. In an appropriate gauge, a general cocycle $\omega\in H^3(\zz_2\times\zz_2,U(1))$ can be written as
\begin{align}\label{eq:Z2Z2_generalcocycle}
  \begin{split}
    \omega(a,b,c) =& \omega_1^{s_1}(a_1,b_1,c_1)\omega_1^{s_2}(a_2,b_2,c_2)\\
    &\times\omega_{II}^{s_3}(a,b,c),
  \end{split}
\end{align}
with $s_i=0,1$ and the group elements are represented by pairs of $\zz_2$ variables, i.e., $a=(a_1,a_2)$. When $s_3=0$, only the cocycles $\omega_1$ appear. They are the same as those seen for the $\zz_2$ phases, defined in Eq.\,\eqref{eq:Z2_cocycle}, only now they depend explicitly on a particular tensor factor, and are referred to as \emph{type-I} cocycles. Cocycles of that type yield TQD models describing a topological order that is decomposable into $\zz_2$ phases. For example, choosing $(s_1,s_2,s_3) = (1,1,0)$ produces a Hamiltonian describing a product of two double semion phases.

In this case, one can make the vertex operators from each copy fully commuting using the same phase modification derived in the previous section. When $s_3=1$, we have a cocycle that can be represented by
\begin{subequations}\label{eq:Z2Z2_typeIIcocycle}
  \begin{align}
      \omega_{II}(a,b,c) =& \omega_1(a_1,b_2,c_2)\\
      \begin{split}
        =& \begin{cases}-1 & a_1=b_2=c_2=1\\ 1 &\text{else}\end{cases},
      \end{split}
  \end{align}
\end{subequations}
which mixes the two tensor factors and therefore is unique to the $\zz_2\times\zz_2$ case.\footnote{Analogously, we could represent this cocycle by $\omega_1(a_2,b_1,c_1)$. However, it is gauge-equivalent to the one we are using \cite{propitius1995topological}.} To distinguish it from the previously studied cocycles, it is referred to as a \emph{type-II} cocycle. A TQD model with such a cocycle as input requires a different modification of the Hamiltonian, which we construct in this section.

The TQD Hamiltonian built by inserting the type-II cocycle from Eq.\,\eqref{eq:Z2Z2_typeIIcocycle} into Eq.\,\eqref{eq:vertex_action} reads
\begin{widetext}
\begin{subequations}\label{eq:Z2Z2_Hamiltonian}
\begin{align}
  \begin{split}
    H_{\zz_2\times\zz_2} =& -\sum_{\text{plaquettes }p}\frac{1}{4}\Big(\mathds{1}
    + \prod_{l\sim p} Z_l^{(1)} \Big)\Big(\mathds{1}
    + \prod_{l\sim p} Z_l^{(2)} \Big)\\
    &- \sum_{\text{vertices }v} \frac{1}{4}\Big( \mathds{1} + A_v^{(1,0)} + A_v^{(0,1)} + A_v^{(1,1)} \Big)
  \end{split}\\
\qq{with} A_v^{(1,0)} =& (-1)^{P_{4^2}^-P_{11^2}^- + P_{6^2}^-P_{12^2}^-}\prod_{l=1}^6X^{(1)}_l,  \\
    A_v^{(0,1)} =& (-1)^{P_{9^1}^-P_{3^2}^- + P_{3^1}^-P_{4^2}^+ + P_{8^1}^-P_{1^2}^- + P_{1^1}^-P_{6^2}^+}\prod_{l=1}^6X^{(2)}_l \qq{and} \\
    A_v^{(1,1)} =& (-1)^{P_{9^1}^-P_{3^2}^- + P_{3^1}^-P_{4^2}^+ + P_{4^2}^+P_{11^2}^- + P_{8^1}^-P_{1^2}^- + P_{1^1}^-P_{6^2}^+ + P_{6^2}^+P_{12^2}^-}\prod_{l=1}^6X^{(1)}_lX^{(2)}_l,
\end{align}
\end{subequations}
\end{widetext}
where $X^{(i)}_l, Z^{(i)}_l$ are the qubit Pauli matrices acting on the $i$th tensor factor. And $P_{l^i}^\pm = \frac{1}{2}(1\pm Z_l^{(i)})$ are the projectors on th values of the $i$th tensor factor of edge $l$.

Again, $\comm*{B_p^{(1)}}{ A_v^g} = \comm*{B_p^{(2)}}{A_v^g} = 0\;\forall g\in\zz_2\times\zz_2$ since each vertex operator flips an even number of qubits around each plaquette. The only obstructions preventing $\{B_p^{(i)}, A_v^g;\; i=1,2;g\in\zz_2\times\zz_2\}$ from forming a pairwise commuting set come from the vertex operators. We will quantify the obstructions below.

\subsubsection{Obstructions in the original model}
Each element in $\zz_2\times\zz_2$ is its own inverse. For the vertex operators to generate a proper stabilizer group, they must represent the group action on site and therefore also square to $\mathds{1}$. Since the representative we chose for the type-II cocycle in Eq.\,\eqref{eq:Z2Z2_typeIIcocycle} does not depend on $a_2$, $b_1$ and $c_1$, we find
\begin{subequations}
\begin{align}
    \left(A_v^{(1,0)}\right)^2 =& \mathds{1}\qq{and}\\
    A_v^{(0,1)}A_v^{(1,0)} =& A_v^{(1,1)}.
\end{align}
\end{subequations}
For the other products of non-trivial group elements $(0,1), (1,0)$ and $(1,1)$ however, we obtain explicit obstructions
\begin{subequations}
\begin{align}
\begin{split}
    \left(A_v^{(0,1)}\right)^2 =& (-1)^{B_{3,9,8,1}^{(1)}}\mathds{1}
    \neq \mathds{1},
\end{split}\label{eq:Z2Z2_obstruction1}\\
\begin{split}
    \left(A_v^{(1,1)}\right)^2 =& (-1)^{B_{3,8,9,1}^{(1)} + B_{6,12,11,4}^{(2)}}\mathds{1}
    \neq \mathds{1},
\end{split}\\
\begin{split}
    A_v^{(1,0)}A_v^{(0,1)} =& (-1)^{B_{6,12,11,4}^{(2)}}A_v^{(1,1)}
    \neq A_v^{(1,1)}
\end{split}
\end{align}
\end{subequations}
where $B_{j,k,l,m}^{(i)} = \frac{1}{2}(\mathds{1} - Z_j^{(i)}Z_k^{(i)}Z_l^{(i)}Z_m^{(i)})$ measures the flux in the $i$th
tensor factor through the region enclosed by the edges $\{j,k,l,m\}$. As in the $\zz_2$ case, the operators square to the identity in the flux-free subspace which includes the ground space.

In addition to the on-site obstructions, the vertex operators fail to commute for neighboring vertices. Due to translation invariance, we need only calculate the commutation relation between $A_v^g$ and the three operators $\{A_i^h,\; i=1,2,3\}$ connected to $v$ by the edges $\{l_1,l_2,l_3\}$ (see Fig.\,\ref{fig:lattice} for the labelling) for any pair $(g,h)$. We obtain the commutativity relations
\begin{subequations}
\begin{align}
    A_{v_2}^{(0,1)}A_v^{(1,0)} =& (-1)^{B_{3,9,8,1}^{(1)}}A_v^{(1,0)}A_{v_2}^{(0,1)},\\
    A_{v_2}^{(1,0)}A_v^{(0,1)} =& (-1)^{B_{3,9,8,1}^{(2)}}A_v^{(0,1)}A_{v_2}^{(1,0)},
\end{align}
\end{subequations}
with the remaining pairs either commuting or not giving independent obstruction phases(see App.\,\ref{app:ZNZN}).
In particular, we find that only the vertex operators acting on different tensor factors fail to commute and they fail precisely when the vertices on which they act are connected by a horizontal edge neighboring a flux. In fact, this is a general property of our model, and can be traced back to the original choice of edge orientations (see App.\,\ref{app:obstructions}).

\subsubsection{Modifying vertex operators by a local phase}
We have found that in the original $\zz_2\times\zz_2$ TQD model the vertex operators fail to be proper stabilizers because, on one hand, the group action is not represented correctly on-site, i.e. $A_v^g A_v^h\neq A_v^{g\oplus h}$ in general, and, on the other, some of them fail to commute. We were able to quantify the obstructions and found that they have a similar structure as in the $\zz_2$ TQD model, namely factors of $-1$ that only depend on fluxes. To resolve the obstructions for the three operators $A_v^{(1,0)}$, $A_v^{(0,1)}$ and $A_v^{(1,1)}$, we modify them by a local phase $D_v^g$ that is the identity on the ground space (Eq.\,\eqref{eq:general_phasemodification}). For the modified operators to be stabilizers, we need them to fulfill
\begin{subequations}\label{eq:Z2Z2_conditions}
\begin{align}
    \left(\Tilde{A}_v^g\right)^2 =& \mathds{1}\quad\forall g\in\zz_2\times\zz_2\label{eq:Z2Z2_condition1},\\
    \Tilde{A}_v^{(1,0)}\Tilde{A}_v^{(0,1)} =& \Tilde{A}_v^{(0,1)}\Tilde{A}_v^{(1,0)} = \Tilde{A}_v^{(1,1)} \label{eq:Z2Z2_condition2},\\
    \comm{\Tilde{A}_v^g}{\Tilde{A}_{v'}^h} =& 0\qcomma v\neq v', \forall g,h.\label{eq:Z2Z2_condition3}
\end{align}
\end{subequations}
The first two conditions are on-site conditions reflecting that the vertex operators should form a representation of the Abelian input group and the third condition is the commutativity condition necessary for the vertex operators to be stabilizers.

\begin{figure*}
    \centering
    \includegraphics[width=0.9\textwidth]{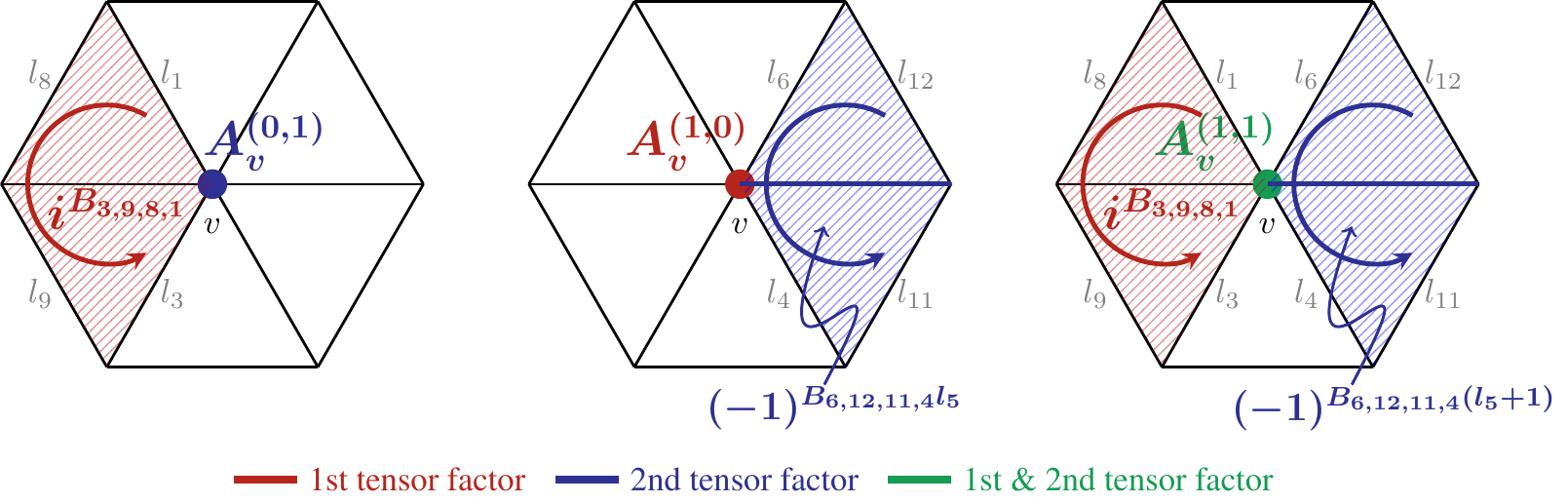}
    \caption{Graphical representation of the stabilizers $\{\Tilde{A}_v^{(0,1)}, \Tilde{A}_v^{(1,0)}, \Tilde{A}_v^{(1,1)}\}$ (left to right). They are composed of TQD vertex operators (Eqs.\,\eqref{eq:Z2Z2_Hamiltonian}) -- represented by the filled dots on the vertex -- and modification phases. The latter are only supported on a restricted neighborhood which is colored red (blue) when acting on the first (second) tensor factor of the local $\mathbb{C}^2\otimes\mathbb{C}^2$ Hilbert space. The edges in gray only enter implicitly via the fluxes. }
    \label{fig:Z2Z2_correctionphases}
\end{figure*}

Condition \eqref{eq:Z2Z2_condition1} gives us independent constraints on each of the \textit{generating phases} $D_v^{(0,1)}$ and $D_v^{(1,0)}$. Using Eq.\,\eqref{eq:Z2Z2_obstruction1}, we find that the first phase must satisfy
\begin{align}
  \begin{split}
    1 =& D_v^{(0,1)} A_v^{(0,1)}D_v^{(0,1)} \left(A_v^{(0,1)} \right)^{-1}\\
    &\times(-1)^{B_{3,9,8,1}^{(1)}},
  \end{split}
\end{align}
which is solved by any solution of the form
\begin{align}
    D_v^{(0,1)} = i^{B_{3,9,8,1}^{(1)}}\bar{D}_v^{(0,1)}
\end{align}
together with $\bar{D}_v^{(0,1)} A_v^{(0,1)}\bar{D}_v^{(0,1)} \big(A_v^{(0,1)} \big)^{-1} = 1$.
Since $A_v^{(1,0)}$ squares to $\mathds{1}$ already, the condition on $D_v^{(1,0)}$ is simpler. Eq.\,\eqref{eq:Z2Z2_condition1} imposes that $1 = D_v^{(1,0)} A_v^{(1,0)}D_v^{(1,0)} \big(A_v^{(1,0)} \big)^{-1}$.  Now, using Eq.\,\eqref{eq:Z2Z2_condition2} to express $D_v^{(1,1)}$ in terms of $D_v^{(1,0)}$ and $D_v^{(0,1)}$ and inserting it into the condition $\big(\Tilde{A}_v^{(1,1)}\big)^2 = \mathds{1}$ yields an additional constraint on $\bar{D}_v^{(0,1)}$ and $D_v^{(1,0)}$,
\begin{align}
\begin{split}
    1 =& A_v^{(0,1)} \bar{D}_v^{(0,1)} \left(A_v^{(0,1)}\right)^{-1}
    A_v^{(1,0)}\bar{D}_v^{(0,1)} \left(A_v^{(1,0)}\right)^{-1}\\
    \times& A_v^{(1,1)}D_v^{(1,0)} \left(A_v^{(1,1)}\right)^{-1} D_v^{(1,0)} (-1)^{ B_{6,12,11,4}^{(2)}}.
\end{split}
\end{align}
A close inspection of this equation shows that this equation is satisfied by
\begin{subequations}
\begin{align}
    \bar{D}_v^{(0,1)} =& 1\qq{and}\\
    D_v^{(1,0)} =& (-1)^{L_5^{(2)}B_{6,12,11,4}^{(2)}},
\end{align}
\end{subequations}
where $L_j^{(i)} = \frac{1}{2}(1-Z_5^{(i)})$ measures the value of the $i$th tensor factor on edge $l_j$ giving rise to the generating phases
\begin{subequations}\label{eq:Z2Z2_generatingphases}
\begin{align}
    D_v^{(0,1)} =& i^{B_{3,9,8,1}^{(1)}},\\
    D_v^{(1,0)} =& (-1)^{L_5^{(2)}B_{6,12,11,4}^{(2)}}
\end{align}
\end{subequations}
The remaining modification phase can be calculated using Eq.\,\eqref{eq:Z2Z2_condition2} to obtain
\begin{align}\label{eq:Z2Z2_11phase}
    D_v^{(1,1)} =& i^{B_{3,9,8,1}^{(1)}}(-1)^{(L_5^{(2)} + 1)B_{6,12,11,4}^{(2)}}.
\end{align}
These solutions define the modified operators
\begin{subequations}
\begin{align}
    \Tilde{A}_v^{(0,1)} =& i^{B_{3,9,8,1}^{(1)}}A_v^{(0,1)},\\
    \Tilde{A}_v^{(1,0)} =& (-1)^{L_5^{(2)}B_{6,12,11,4}^{(2)}}A_v^{(1,0)}\qq{and}\\
    \Tilde{A}_v^{(1,1)} =& i^{B_{3,9,8,1}^{(1)}} (-1)^{(L_5^{(2)} + 1)B_{6,12,11,4}^{(2)}} A_v^{(1,1)},
\end{align}
\end{subequations}
which form a faithful representation of $\zz_2 \times \zz_2$. When inserting these operators into Eq.\,\eqref{eq:Z2Z2_condition3}, one finds that they also commute for neighboring vertices which completes our construction of non-Pauli stabilizers $\{\Tilde{A}_v^{(1,0)},\Tilde{A}_v^{(0,1)},\Tilde{A}_v^{(1,1)}\}$ based on the $\zz_2\times\zz_2$ TQD model constructed with a type-II cocycle. Note that -- unlike in the $\zz_2$ case -- we have not found a one-parameter family of solutions, though we do not claim that our solution is unique. Thankfully, the modification phases we derived here are already quite simple in form, depending only on a restricted neighborhood of the vertex, illustrated in Fig.\,\ref{fig:Z2Z2_correctionphases}. The modification phase of $A_v^{(0,1)}$ adds an $i$ whenever $B_{3,9,8,1}^{(1)}=1 \mod 2$, which can be seen as an $S$-gate on the flux. Similarly, one can see that the phase for $A_v^{(1,0)}$, $(-1)^{B_{6,12,11,4}^{(2)}L_5^{(2)}}$, is a controlled $Z$-gate between the flux $B_{6,12,11,4}$ and the edge $l_5$.

With these modifications, we have constructed a set of stabilizers whose code space is the ground space of a $\zz_2\times\zz_2$ topological order that cannot be \emph{factored} into two (possibly twisted) $\zz_2$ phases. It is just one example of how our analysis of the on-site and commutativity obstructions in a TQD model allows us to obtain stabilizers from various topological orders, since the techniques used here can be extended to more general models.

\subsection{General Abelian topological order}
In the previous subsections, we explicitly calculated and corrected the obstructions to the construction of stabilizers from qubit-based TQD models. The method can also be applied for the input groups $\zz_N=\{0,1,\dots,N-1\}$ and $\zz_N\times\zz_N$ with type-I and type-II cocycles (and combinations thereof). For those topological orders, the construction of the modification phases follows a similar line as that of the qubit based models. In the following, we will sketch the general construction and state the resulting modification phases for $\zz_N$ and $\zz_N\times\zz_N$. The detailed calculation can be found in App.\,\ref{app:ZN} and \ref{app:ZNZN}.

The action of the vertex operator in terms of cocycles (Eq.\,\eqref{eq:vertex_action}) allows us to quantify the on-site and the commutativity obstructions for a generic TQD model.
Moreover, we can derive consistency equations for the modification phase by imposing that the vertex operators should represent the group action on-site and commute for neighboring vertices (see App.\,\ref{app:obstructions}).
Using the canonical representative of a type-I cocycle for $\zz_N$,
\begin{align}\label{eq:ZNcocycle}
    \omega_I(a,b,c) = e^{\frac{2\pi i}{N^2}a(b+c-[b+c]_N)},
\end{align}
where $[a+b]_N = (a+b)\mod N$, allows us to explicitly solve the consistency equations and obtain pairwise commuting vertex operators $\{\Tilde{A}_v^g\}$ that represent the group action of $\zz_N$. We exploit the cyclicity of $\zz_N$ by imposing that every vertex operator to the $N$th power should equal the identity, just as in Eq.\,\eqref{eq:Z2_condition1} for $N=2$. In particular, this should hold for the \textit{generating vertex operator} $A_v^1$, which allows us to determine a suitable ansatz for the corresponding modification phase $D_v^1$. From this, we find a family of solutions $D_v^1(p)$ that ensure that the generating vertex operators not only represent the group but also commute pairwise. The fact that $\Tilde{A}_v^1$ generates every other $\Tilde{A}_v^g$ allows us to compute every other modification phase $D_v^g$ iteratively. One of the resulting modification phases for any $g\in\zz_N$ reads
\begin{align*}
    D_v^g =& e^{\frac{2\pi i}{N^2} gB_{6,12,11,4}}e^{-\frac{2\pi i}{N} g\sum_{n=0}^{-L_2-g} \left( P_{-2,8,1}^{(n)} - P_{-2,9,3}^{(n)}\right)}\\
  \times& e^{-\frac{2\pi i}{N} \sum_{n=0}^{g-1}n\left(  P_{-2,8,1}^{(-L_2-n)} - P_{-2,9,3}^{(-L_2-n)}\right)} \numberthis\label{eq:ZNcorrectionphase}\\
  \times& e^{-\frac{2\pi i}{N} \sum_{n=0}^{g-1}n \left(P_{6,12,-5}^{(-L_5+n)} - P_{4,11,-5}^{(-L_5+n)}\right)},
\end{align*}
where $P_{i,j,k}^{(n)}$ is the projector onto the space in which the sum of edge values $l_i + l_j + l_k = n \mod N$ and a minus sign in front of an index states that the inverse element enters in the sum. For example, the projector $P_{-2,8,1}^{(n)}$ projects onto the space in which $-l_2+l_8+l_1 = n\mod N$. The flux $B_{6,12,11,4} = (L_6 + L_{12} - L_{11} - L_4) \mod N$ is defined in a similar fashion as in the $\zz_2$ case. Unlike the $\zz_2$ case, we have to take the orientation of the edges into account by subtracting $l_{11}$ and $l_4$ since the elements of $\zz_N$ are not self-inverses.
We note that the modification phase above is 1 in the zero-flux subspace, where $B_{6,12,11,4}=0$ and all the projectors appearing cancel pairwise in the above expression. The first term is a $N^2$th root of unity and only depends on fluxes. It is the higher dimensional analogue of the first factor in the $\zz_2$ solution, Eq.\,\eqref{eq:Z2_correctionphasep1}. The second term, which only includes the edge value $l_2$ in the upper summation bound, reduces to the second term in Eq.\,\eqref{eq:Z2_correctionphasep1} when $N=2$. The only term where $l_2$ enters in the argument of the projectors is the third term. For $N=2$, this term reduces to the last term in Eq.\,\eqref{eq:Z2_correctionphasep1}.
For a detailed derivation of the modification phase for $G=\zz_N$ we refer to App.\,\ref{app:obstructions} and \ref{app:ZN}.

When considering the TQD model with gauge group $\zz_N^2$ and a type-II cocycle (Eq.\,\eqref{eq:Z2Z2_typeIIcocycle}), the construction follows a similar path. For each tensor factor, the vertex operators must fulfill the same closure relation as in $\zz_N$. This allows us to find suitable ansatzes for the modification phases $D_v^{(0,1)}$ and $D_v^{(1,0)}$ for the two generating vertex operators $\Tilde{A}_v^{(0,1)}$ and $\Tilde{A}_v^{(1,0)}$ so that the modified operators $\{\Tilde{A}_v^{(0,1)}, \Tilde{A}_v^{(0,1)}\}$ represent the group action of the two generators $(0,1)$ and $(1,0)$ on-site in a consistent fashion on both tensor factors and commute pairwise. Since they correspond to the two generators of $\zz_N\times\zz_N$, we can again iteratively construct the modification phase for any $g = (g_1,g_2)\in\zz_N^2$,
\begin{align*}
    D_v^{(g_1,g_2)}& = e^{\frac{2\pi i}{N^2} g_2 B_{3,9,8,1}^{(1)}} \numberthis\\
   &\times e^{-\frac{2\pi i}{N}g_1 \sum_{i=0}^{-L_5^{(2)}-1+g_2}\big(P_{(6,12,-5)^{(2)}}^{(i)}-P_{(4,11,-5)^{(2)}}^{(i)}\big)},
\end{align*}
where we indicate the operators only supported on the $i$th tensor factor with an upper index in brackets, e.g. $L_5^{(1)}$ acts like $L_5\otimes \mathds{1}$ on the $\mathbb{C}^N\otimes\mathbb{C}^N$ Hilbert space on edge 5.
The flux $B_{3,9,8,1}^{(i)} = (-L_3^{(i)} - L_9^{(i)} + L_8^{(i)} + L_1^{(i)}) \mod N$ measures the $i$th tensor factor of the flux through the diamond left of the vertex (see. Fig.\,\ref{fig:Z2Z2_correctionphases}) and the projectors $P_{(i,j,k)^{(l)}}^{(n)}$ are defined as in the $\zz_N$ case but on the $l$th tensor factor. As with the $\zz_N$ solution, we can identify the terms derived in the previous subsection for $N=2$ (see Eqs.\,\eqref{eq:Z2Z2_11phase} and \eqref{eq:Z2Z2_generatingphases}). The first term only depends on fluxes and is non-trivial when $g_2\neq 0$. In contrast, the second term depends on both tensor factors of $g$, where $g_1$ defines which $N$th root of unity is appended and $g_2$ selects which projectors appear in the sum. The explicit edge value $l_5^{(2)}$ enters only in the upper summation bound of the second term and has a similar influence on the sum of projectors as $g_2$. In contrast to the one-parameter family of the $\zz_N$ case, we have found only a single solution. For a detailed derivation, see App.\,\ref{app:ZNZN}.

Although we have only calculated the modification phase for $\zz_N$ and $\zz_N\times\zz_N$, these results readily generalize to any \emph{Abelian group} yielding an \emph{Abelian anyon} theory. Since, by the fundamental theorem of finitely generated Abelian groups, any finite Abelian group can be decomposed into $\zz_N$ factors. As a consequence of this, one can also decompose the cocycles of an Abelian group into cocycles of of this cyclic decomposition\cite{propitius1995topological}. Whilst there exist cocycles beyond the type-I and type-II classes considered here, their inclusion in a TQD model produces topological order containing non-Abelian anyons, and are therefore not suitable for stabilizer error correction based on commuting syndrome measurements. In our construction, we have explicitly constructed the modification phase for type-I and type-II cocycles. Since the cocycle defining an arbitrary Abelian TQD model will be a product of type-I and type-II cocycles, the modification phases can be computed as above for each factor in that product. The resulting phases can then be multiplied together as they are all diagonal operators and therefore commute. This finalizes the argument that our construction carries over to any Abelian topological order derived from a TQD model.

\subsection{Towards quantum error correction}

In this subsection, we discuss the potential of the new
topological codes devised here for notions of
fault tolerant quantum computing and
quantum error correction \cite{RevModPhys.87.307}.
We have constructed pairwise commuting operators $\Tilde{A}_v^g$ that, together with $B_p^g$ (see Eq.\,\eqref{eq:Bp_decomposition}) generate the Non-Pauli stabilizer group $\mathcal{S}_{TQD}$. Just as in other topological codes, a measurement of all generators of this group gives rise to a unique excitation pattern. This diagnostic information can be used
in \emph{decoders} that allow for an eventual correction
by annihilating them with suitable string-like operators. The same string operators, when closed around a non-trivial loop, give rise to logical operators of the codes.

\begin{figure}
  \centering
  \includegraphics[width=0.45\textwidth]{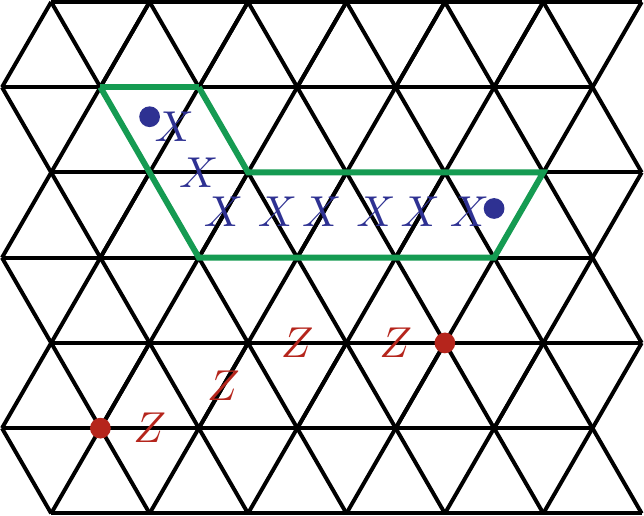}
  \caption{Example of two elementary string operators of the double semion code. A $Z$-string along a path on the lattice creates vertex excitations at its endpoints (red dots). Similarly, a pair of plaquette excitations (blue dots) is created by a $X$-string along a path on the dual lattice which is dressed with a phase factor $\beta_{\mathcal{P}}$. The latter is supported on the neighborhood of the path, whose boundary is depicted in green. The logical operators of the code are generated by these string operators acting on along non-trivial loops on the surface.}
  \label{fig:stringoperators_exmpl}
\end{figure}

Let us illustrate the qualitative difference of the string operators in our codes to the toric code-like string operators in Kitaev's \emph{quantum double models} \cite{kitaev2003fault} with the previously discussed TQD model with $G=\zz_N$ and a non-trivial cocycle from Eq.\,\eqref{eq:ZNcocycle}. The group action can be implemented by a generalized Pauli operator $X: \ket{n}\mapsto\ket{n+1 \mod N}$. With that, the vertex operators are of the form
\begin{align}
  A_v^g = \alpha_v^g \prod_{l\sim v} X^{gs(l,v)}_l,
\end{align}
where $\alpha_v^g$ is a (in the computational basis) diagonal unitary operator containing the pre-factor of the original vertex operator made up of cocycles (see Eq.\,\eqref{eq:vertex_action}) and the correction phase $D_v^g$ (see Eq.\,\eqref{eq:ZNcorrectionphase}). The sign $s(l,v)=(-1)+1$ for an (outgoing)incoming edge $l$ w.r.t.\ vertex $v$ (compare Eqs.\,\eqref{eq:vertex_action_edges}).
The plaquette operators are the known plaquette operators $B_p^g$ made up of generalized Pauli $Z$s around the plaquette. The elementary string operators of the corresponding code have the property that they are supported on paths on the (dual) lattice and commute everywhere with the stabilizers except at their endpoints where they only commute up to a phase $e^{\pm 2\pi i/N}$ with either the plaquette or vertex operators. Since the plaquette operators are exactly the same as in the \emph{($N$-level) toric code}, the string operators that create a pair of vertex-excitations (charges) are strings of Pauli $Z$s along a path connecting the two vertices that carry the excitations. The elementary string operators supported along a (oriented) path $\mathcal{P}$ in the dual lattice creating fluxes at their end-plaquettes, on the other hand, have to be of the form
\begin{align}
  S^g_{\mathcal{P}} = \beta_{\mathcal{P}}^g \prod_{l\in\mathcal{P}}X_l^{g s(l,\mathcal{P})},
\end{align}
where $\beta_{\mathcal{P}}^g$ is a diagonal operator similar to $\alpha_v^g$ but supported on the path $\mathcal{P}$ and its nearest neighbors and $s(l,\mathcal{P})=(-1)1$ for an edge $l$ that crosses $\mathcal{P}$ from the (left)right. The reason for that is that a simple $X$ string on the dual lattice does not commute with the vertex operators due to the non-trivial pre-factor $\alpha_v^g$. For the double semion code, the string operators are depicted in Fig.\,\ref{fig:stringoperators_exmpl}.

To find the pre-factors for the string operators $\beta_{\mathcal{P}}^g$ it suffices to look at the generator $g=1$ and then define $S_\mathcal{P}^g=\left(S_\mathcal{P}^1\right)^g$. Moreover, strings along longer paths can be formed by concatenation of elementary strings along paths of length 1. This allows us to generate any flux-string operator for the code (and with that the logical operators as well) from \textit{generating string operators} that create excitations on neighboring plaquettes and then use the same operators to move one excitation away from the other.
These generating strings can be obtained as solutions to the consistency conditions
\begin{subequations}
  \begin{align}
    \left(S^1_{\mathcal{P}}\right)^N =& \mathds{1},  \\
    \comm{S^1_{\mathcal{P}}}{A_v^1}_G =& \mathds{1} \qq{}\forall v,
  \end{align}
\end{subequations}
where $\comm{A}{B}_G=A B A^{-1}B^{-1}$ denotes the group commutator and $N$ is the order of the corresponding excitation.
The first condition enforces the correct fusion rules of the underlying excitations and the second one ensures that $S_\mathcal{P}^1$ is indeed an elementary string operator of the code as discussed above.
The authors of Ref.\ \cite{dauphinais2019quantum}
Ref.\ have encountered similar conditions in their version of a semion code ($N=2$ in our case) and have identified
an algorithmic way to construct the phase $\beta_v^1$. This method should be generalizable to higher local dimensions $N>2$ and models from product groups like the \emph{twisted color codes} introduced above. This -- as well as identifying
ways of making the best
use of the diagnostic information obtained in non-Pauli stabilizer measurements -- will be the
topic of future work.

What all TQD codes from a non-trivial cocycles have in common is the exotic structure of the vertex operators and the flux-string operators, as illustrated above. This results in strings of (generalized) Pauli $X$ errors no longer only creating flux-excitations at their endpoints but also creating correlated vertex excitations in its interior. In particular, such a Pauli $X_{\mathcal{P}}$ string along a path $\mathcal{P}$ will decompose into elementary string operators as
\begin{align}
  X_{\mathcal{P}} = \sum_{\{P_z\}} c_{\mathcal{P}}(P_z) P_z S_{\mathcal{P}}^1,
\end{align}
where $\{P_z\}$ is the set of possible products of (generalized) Pauli $Z$s supported on the neighborhood of path $\mathcal{P}$ and $\{c_{\mathcal{P}}(P_z)\}$ are the (complex) expansion coefficients that can be obtained by projecting $\beta_\mathcal{P}^1$ onto the Pauli words $\{P_z\}$. This means that Pauli errors act on the TQD codes in a similar way as coherent errors in Pauli stabilizer codes. Investigating TQD codes with Pauli noise therefore could also bring insight into the study of coherent errors in Pauli codes.

\section{Conclusion and outlook}\label{sec:conclusion}
In this work, we have exploited the deep connections between topological phases of matter and topological error correction to construct a new class of stabilizer codes built from twisted quantum double models hosting Abelian anyons. To do so, we have established a systematic and quantitative understanding of how the vertex operators of twisted quantum double models fail to commute outside of the ground space and therefore precluding their use as stabilizers without further modification. We began with the relatively straightforward task of deriving the obstructions for the fixed-point Hamiltonian of the double semion phase -- the twisted version of the toric code Hamiltonian -- and of a twisted $\zz_2\times\zz_2$ phase. By appropriately modifying the vertex operators, we have obtained commuting stabilizers from both models that, in principle, can be implemented with a two dimensional qubit architecture. This approach readily generalizes to other twisted quantum double models with higher local dimensions. We have explicitly constructed commuting stabilizers from the twisted $\zz_N$ and $\zz_N^2$ models, making it possible to construct stabilizers for \emph{every} Abelian TQD model.

Our findings invite further research into explicit error correction schemes based on these novel stabilizers so that their potential may be fully explored. For any decoder proposed, one has to find suitable string operators for these modified models, which provide both the recovery operations and the logical operations on the code space. We expect that the feature of our codes that Pauli $X$ strings (and their higher dimensional analogues) not only create plaquette excitations at their endpoints but also some vertex excitations along their path can be used to design a tailored decoder with potentially increased performance under Pauli noise. A first step in this direction has been taken in Ref.\
\cite{varona2020determination}
 using a neural network decoder for their \emph{semion code}. Although they did not find substantial evidence of increased performance compared to the \emph{toric code}, it is far from clear whether there does not exist a better decoder tailored to the exotic string operators and what can be achieved with Stabilizer codes implementing other topological orders -- like the \emph{twisted color code} introduced in Sec.\,\ref{sec:Z2Z2}. Moreover, a more sophisticated analysis of the behaviour of our codes under Pauli noise would not only yield a deeper understanding of Non-Pauli topological codes in general but also connect to Pauli (topological) codes with coherent errors: In fact, here one has to deal with similar a decomposition of errors into elementary strings, but in reversed roles. As far as we know, mappings to statistical mechanics models for such error models
 \cite{Chubb}, for example, are yet to be discovered.

On a more abstract level, our codes host fundamentally different anyons to those in Pauli stabilizer codes resulting in a different algebra of the logical operators, twist defects and domain walls \cite{kesselring2018boundaries} and with that allow for a different fault-tolerant gate set. With our stabilizers at hand, it is now possible to implement these \textit{twisted} anyon theories in an error correcting code inviting research into their potential for topological quantum computation.
Though these codes in and of themselves are a novel take on the idea of stabilizer-based error correction, it is our hope that the many questions raised by this class of codes spurs the comprehensive
investigation into their properties and their
potential for fault-tolerant quantum computing.


\section*{Acknowledgements}
This work has been supported by the DFG (CRC 183, EI 519/15-1), and
the Templeton Foundation. We thank B. J.\ Brown, M.\ Kesselring, C.\ Wille, J.\ Conrad,
and J.\ Haferkamp for fruitful discussions and the JARA Institute for Quantum Information at RWTH Aachen for its hospitality during a large time period of this research. Moreover, we thank F. Hassler for supporting this collaboration.

\bibliographystyle{plainnat}
\bibliography{references}

\onecolumn
\newpage
\appendix
\section{Group cohomology with $U(1)$ coefficients}\label{app:group_cohomology}
In this section, we will give an algebraic definition of group cohomology with $U(1)$ coefficients. In some sense it can be thought of a condensed version of the appendix in Ref.~\cite{chen2013symmetry} in which we give
all the necessary background to understand our general framework described in the next appendices. Besides that, it is an interesting subject on its own and pops up in different fields of physics and mathematics.

In our context, we deal with the cohomology of groups over $U(1)$.\footnote{Formally, here $U(1)$ is a $G$-module with trivial group action.} To define the $n$th cohomology group, we start by defining maps from multiple copies of $G$ onto $U(1)$, $\omega_n: G^n \to U(1)$. Such a general map is called $n$-\textit{cochain} and we denote the set of all such functions by $\mathcal{C}^n(G,U(1))$. Moreover, we define the so called \textit{coboundary} operator mapping $n$-cochains to $n+1$-cochains, $\delta_n: \mathcal{C}^n(G,U(1)) \to \mathcal{C}^{n+1}(G,U(1))$ with
\begin{align}
  \begin{split}
    (\delta_n \omega_n)(g_0,g_1,\dots,g_n) =& \omega_n^{s(g_0)}(g_1,\dots,g_n)\prod_{i=0}^{n-1} \omega_n^{(-1)^{i+1}}(g_0,\dots,g_{i-1},g_ig_{i+1},\dots,g_n)\\
    &\times\omega_n^{(-1)^{n+1}}(g_0,\dots,g_{n-1}),
  \end{split}
\end{align}
where $s(g)=-1$ if $g$ is antiunitary and 1 if it is unitary and the group multiplication symbol on $G$ between $g_i$ and $g_{i+1}$ is implicit. A short calculation shows that $(\delta_{n+1}\circ\delta_{n}) \omega_n = 1\;\forall \omega_n\in\mathcal{C}^n(G,U(1))$ which is the defining property of any coboundary operator.
For $n=2$ and $n=3$ for example the coboundary operator acts as
\begin{subequations}
\begin{align}
    (\delta_2\omega_2)(g_0,g_1,g_2) =& \frac{\omega_2(g_1,g_2)\omega_2(g_0,g_1g_2)}{\omega_2(g_0g_1,g_2)\omega_2(g_0,g_1)}\label{eq:app_2coboundary},\\
\begin{split}
    (\delta_3\omega_3)(g_0,g_1,g_2,g_3) =&
    \frac{\omega_3(g_1,g_2,g_3)\omega_3(g_0,g_1g_2,g_3)\omega_3(g_0,g_1,g_2)}{\omega_3(g_0g_1,g_2,g_3)\omega_3(g_0,g_1,g_2g_3)}.\label{eq:app_3coboundary}
\end{split}
\end{align}
\end{subequations}
The first thing we do with an algebraic map is to define its kernel and its image. We call any element in the kernel of $\delta_n$ \textit{$n$-cocycle} and denote the set of $n$-cocycles as $\mathcal{Z}^n(G,U(1)) = \{\omega\in\mathcal{C}^n(G,U(1));\; \delta_n\omega = 1\}$. We call any element in the image of $\delta_{n-1}$ \textit{n-coboundary} and denote the set of $n$-coboundaries by $\mathcal{B}^n(G,U(1)) = \{\omega\in\mathcal{C}^n(G,U(1));\; \omega=\delta_{n-1}\beta,\beta\in\mathcal{C}^{n-1}(G,U(1))\}$.

Due to $(\delta_n\circ\delta_{n-1}) \omega_{n-1} = 1$ we can multiply any $n$-cocycle with an $n$-coboundary and the result will still be a $n$-cocycle. Group cohomology classifies all the inequivalent cocycles under such a \textit{gauge freedom}, i.e. the $n$th cohomology group of $G$ over $U(1)$ is defined by
\begin{align}
    H^n(G,U(1)) = \faktor{\mathcal{Z}^n(G,U(1))}{\mathcal{B}^n(G,U(1))}.
\end{align}
By comparison of Eqs.\,\eqref{eq:cocycle_condition} and \eqref{eq:coboundary_gauge} with Eqs.\,\eqref{eq:app_2coboundary} and \eqref{eq:app_3coboundary}, we see that the classification of the topological orders of a twisted gauge theory with gauge group $G$ in Eq.\,\eqref{eq:formalclassification} is exactly what we defined here as the third cohomology group of $G$ over $U(1)$.

\section{Obstruction in general TQD models from Abelian groups}\label{app:obstructions}
Consider a finite Abelian group $G$ and a 3-cocycle $\omega\in H^3(G,U(1))$. By construction, the vertex operators $\{A_v^g\}$ in the original TQD model defined by $(G,\omega)$ (see Eq.\,\eqref{eq:vertex_action}) form a representation of the group on the flux-free Hilbert space, where $B_p = 1\;\forall p$, i.e., they implement the group action on-site and commute with any vertex operator acting on a different vertex. In the following, we will investigate the how they act on the total Hilbert space $\mathcal{H}=\bigotimes \mathcal{H}_l$. In particular, we find two types of obstructions -- one which capture the failure of the group multiplication rule when implemented on site (the \emph{on-site obstruction}) and one which captures the failure of two neighboring operators to commute (the \emph{commutativity obstruction}).

\subsection{On-site obstruction}
We will first investigate the successive action of two vertex operators $A_v^g$ and $A_v^h$ on an arbitrary basis element. Using the action defined in Eq.\,\eqref{eq:vertex_action}, we obtain
\begin{align}
    \begin{split}
  A_v^gA_v^h \Bigg| \begin{minipage}[c]{2.2cm}
    \includegraphics[width=\textwidth]{basis_config}
  \end{minipage}
 \Bigg\rangle =& \frac{\omega(l_9,l_3 h^{-1},h)\omega(l_3 h^{-1},h,l_4)\omega(h,l_4,l_{11})}{\omega(l_8,l_1 h^{-1},h)\omega(l_1 h^{-1},h,l_6)\omega(h,l_6,l_{12})}\\
&\times \frac{\omega(l_9,l_3'',g)\omega(l_3'',g,h l_4)\omega(g,h l_4,l_{11})}{\omega(l_8,l_1'',g)\omega(l_1'',g,h l_6)\omega(g,h l_6,l_{12})} \Bigg|\begin{minipage}[c]{2.2cm}
   \includegraphics[width=\textwidth]{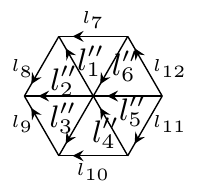}
 \end{minipage}\Bigg\rangle,
   \end{split}
\end{align}
where $l_i'' = l_i h^{-1}  g^{-1}$ for $i=1,2,3$ and $l_j'' = g h l_j$ for $j=4,5,6$. Note that we have left the group multiplication symbol implicit. Besides shifting the edge values of $l_1$ to $l_6$, the two vertex operators multiply the basis vector with a phase factor given by a product of 12 cocycles. We can simplify this large product using the cocycle condition Eq.\,\eqref{eq:cocycle_condition} so that it reduces to a product of 10 cocycles
\begin{align}
\begin{split}
    \frac{\omega(l_9 l_3'',g,h)\omega(g,h,l_4 l_{11}) \omega(l_9,l_3'',g h)\omega(l_3'',g h,l_4)\omega(g h,l_4,l_{11})}{\omega(l_8 l_1'',g,h)\omega(g,h,l_6 l_{12}) \omega(l_8,l_1'',g h)\omega(l_1'',g h,l_6)\omega(g h,l_6,l_{12})},
\end{split}
\end{align}
where we can identify the pre-factor of $A_v^{g h}$
\begin{align}
    A_v^gA_v^h \Bigg| \begin{minipage}[c]{2.3cm}
    \includegraphics{basis_config}
  \end{minipage}
 \Bigg\rangle = \frac{\omega(l_9 l_3'',g,h)\omega(g,h,l_4 l_{11})}{\omega(l_8 l_1'',g,h)\omega(g,h,l_6 l_{12})} A_v^{g h} \Bigg|\begin{minipage}[c]{2.3cm}
   \includegraphics{basis_config}
 \end{minipage}
 \Bigg\rangle.\label{eq:app_grouppbstruction_basis}
\end{align}
Since this equality holds for any basis element and any group elements $g,h$, we have established a relation between the operators $A_v^gA_v^h$ and $A_v^{g h}$. By rewriting the product of edges in terms of the fluxes $b_{3,9,2} = l_3^{-1} l_9^{-1}  l_2$, $b_{2,8,1} = l_2^{-1} l_8 l_9$, $b_{5,11,4} = l_5 l_{11}^{-1} l_4^{-1}$ and $b_{6,12,5} = l_6 l_{12} l_5^{-1}$, this relation reads
\begin{align}
\begin{split}
    A_v^gA_v^h =& \frac{\omega(l_2 (b_{3,9,2})^{-1},g,h)\omega(g,h,(b_{5,11,4})^{-1} l_5  h^{-1} g^{-1})}{\omega(l_2 b_{2,8,1},g,h)\omega(g,h,b_{6,12,5} l_5  h^{-1} g^{-1})} A_v^{g h}\\
  \eqqcolon& \Omega_{\{b\}}^{(l_2,l_5)}(g,h) A_v^{g h},
\end{split}\label{eq:app_onsiteobstruction}
\end{align}
where we have omitted the group multiplication symbol for clarity and introduced the \textit{on-site obstruction phase} $\Omega_{\{b\}}^{(l_2,l_5)}(g,h)$, defined with a fixed flux configuration $\{b\}$, that is only supported on the horizontal edges $l_2$ and $l_5$ and depends on the two group elements $g,h$.

By introducing the fluxes as above, we immediately see that the on-site obstruction phase equals 1 in the flux-free subspace. Hence, $A_v^gA_v^h = A_v^{gh}$ holds in the absence of fluxes, as anticipated. On the whole Hilbert space however, group multiplication is only faithfully implemented up to the phase factor $\Omega$.\footnote{One might be eager to see the vertex operators as a projective representation of $G$. However, one has to be
careful with this, since $\Omega$ is in fact an operator that itself does not commute with the vertex operators. Hence, where one might think only of projective representations characterized by group cohomology over a module with trivial action as discussed in the previous section, one really is dealing with a group cohomology over a module with non-trivial action characterizing the obstruction phase, sometimes referred to as twisted group cohomology.}

\subsection{Commutativity obstruction}
To quantify the (non-)commutativity of the vertex operators in the original TQD model, we compute the group commutator $\comm{A_{v'}^g}{A_v^h}_G\coloneqq A_{v'}^gA_v^h(A_{v'}^g)^{-1}(A_v^h)^{-1}$. It is clear from the definition that two vertex operators that act on vertices separated by two or more edges commute. We therefore only have to consider neighboring vertices $v$ and $v'$.
Due to the translation symmetry of our lattice, we only need to consider three cases: Whether $v$ and $v'$ are connected by $l_1$, $l_2$ or $l_3$ (see Fig.\,\ref{fig:lattice}).

In the first case, adapting the labelling of the vertices in Fig.\,\ref{fig:lattice} and using Eq.\,\eqref{eq:vertex_action}, the group commutator, acting on an arbitrary basis vector, reads
\begin{align}
\begin{split}
    \comm{A_1^g}{A_v^h}_G \Bigg|   \begin{minipage}[c]{1.4cm}\includegraphics[width=\textwidth]{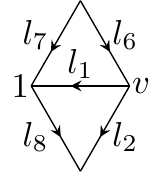} \end{minipage}%
  \Bigg\rangle =&
  \frac{\omega(g,g^{-1}l_1,l_6) \omega(l_8,g,g^{-1}l_1)}{\omega(l_8g,g^{-1}l_1,h) \omega(g^{-1}l_1,h,h^{-1}l_6)}\\ &\times\frac{\omega(l_8, l_1,h) \omega(l_1,h,h^{-1}l_6)}%
  {\omega(l_8,g,g^{-1}l_1 h) \omega(g,g^{-1}l_1 h, h^{-1}l_6)} \Bigg|   \begin{minipage}[c]{1.4cm}\includegraphics[width=\textwidth]{diamond2} \end{minipage}%
  \Bigg\rangle,
\end{split}
\end{align}
where we have only explicitly shown the part of the lattice on which both operators act non-trivially. Using the two cocycle conditions (Eq.\,\eqref{eq:cocycle_condition})
\begin{subequations}
\begin{align}
    \frac{\omega(l_8, l_1,h)\omega(l_8,g,g^{-1}l_1)}{\omega(l_8g,g^{-1}l_1,h)\omega(l_8,g,g^{-1}l_1h)} \stackrel{\eqref{eq:cocycle_condition}}{=}& \frac{1}{\omega(g,g^{-1}l_1,h)}\qq{and}\\
    \frac{\omega(g,g^{-1}l_1,l_6)\omega(l_1,h,h^{-1}l_6)}{\omega(g^{-1}l_1,h,h^{-1}l_6)\omega(g,g^{-1}l_1h,h^{-1}l_6)}\stackrel{\eqref{eq:cocycle_condition}}{=}& \omega(g,g^{-1}l_1,h),
\end{align}
\end{subequations}
we see that this pre-factor is in fact equal to $1$ for any cocycle.
Since this holds for any basis vector, $\comm{A_1^g}{A_v^h}_G = \mathds{1}$ on the whole Hilbert space and the vertex operators $A_1^g$ and $A_v^h$ commute in the original TQD model. We observe the same for the vertex operators $A_3^g$ and $A_v^h$, since
\begin{align}
\begin{split}
  \comm{A_3^g}{A_v^h}_G \Bigg|   \begin{minipage}[c]{1.4cm}\includegraphics[width=\textwidth]{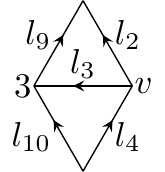} \end{minipage}%
  \Bigg\rangle =&
  \frac{\omega(g,g^{-1}l_1,l_6) \omega(l_8,g,g^{-1}l_1) \omega(l_8, l_1,h) \omega(l_1,h, h^{-1}l_6)}%
  {\omega(l_8g,g^{-1}l_1,h) \omega(g^{-1}l_1,h,h^{-1}l_6) \omega(l_8,g,g^{-1}l_1 h) \omega(g,g^{-1}l_1 h, h^{-1}l_6)} \Bigg|   \begin{minipage}[c]{1.4cm}\includegraphics[width=\textwidth]{diamond3} \end{minipage}%
  \Bigg\rangle\\
  \stackrel{\eqref{eq:cocycle_condition}}{=}& \Bigg|   \begin{minipage}[c]{1.4cm}\includegraphics[width=\textwidth]{diamond3} \end{minipage}%
  \Bigg\rangle,
\end{split}
\end{align}
where we have again used the cocycle condition to identify the pre-factor in the first line with $1$.

This leaves only the second pair of vertices, $v$ and $2$, where we keep to the same procedure. However, when we compute the commutator, Eq.\,\eqref{eq:vertex_action},
\begin{align}
\begin{split}
    \comm{A_2^g}{A_v^h}_G \Bigg| \begin{minipage}[c]{1.4cm}\includegraphics[width=\textwidth]{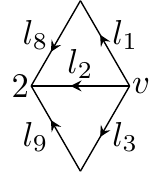} \end{minipage} \Bigg\rangle =
    \frac{\omega(g^{-1}l_9,l_3,h)\omega(g,g^{-1}l_9,l_3)\omega(g,g^{-1}l_8,l_1h)\omega(l_8,l_1,h)}{\omega(g^{-1}l_8,l_1,h)\omega(g,g^{-1}l_8,l_1)\omega(g,g^{-1}l_9,l_3h)\omega(l_9,l_3,h)}\Bigg| \begin{minipage}[c]{1.4cm}\includegraphics[width=\textwidth]{diamond1} \end{minipage} \Bigg\rangle,
\end{split}\label{eq:app_commutativity2}
\end{align}
we find that -- in contrast to the previous cases -- no amount of cocycle manipulation can remove the pre-factor, leaving a phase of the form
\begin{align}
    \frac{\omega(g^{-1}l_9,l_3,h)\omega(g,g^{-1}l_9,l_3)\omega(g,g^{-1}l_8,l_1h)\omega(l_8,l_1,h)}{\omega(g^{-1}l_8,l_1,h)\omega(g,g^{-1}l_8,l_1)\omega(g,g^{-1}l_9,l_3h)\omega(l_9,l_3,h)}= \frac{\omega(g,g^{-1}l_8l_1,h)}{\omega(g,g^{-1}l_9l_3,h)}.
\end{align}
Since Eq.\,\eqref{eq:app_commutativity2} holds for any basis vector we established an identity for the operator $\comm{A_2^g}{A_v^h}_G$. Expressed in terms of the fluxes $b_{2,8,1}$ and $b_{3,9,2}$ introduced in Eq.\,\eqref{eq:app_onsiteobstruction} it reads
\begin{align}
\begin{split}
    \comm{A_2^g}{A_v^h}_G =& \frac{\omega(g,g^{-1}l_2b_{2,8,1},h)}{\omega(g,g^{-1}l_2(b_{3,9,2})^{-1},h)}\eqqcolon \Pi_{\{b\}}^{(l_2)}(g,h),
\end{split}\label{eq:app_commutativityobstruction}
\end{align}
where we introduced the \textit{commutativity obstruction phase} $\Pi_{\{b\}}^{(l_2)}(g,h)$ similar to the on-site obstruction phase $\Omega$ in the previous section. Note that, for a given flux configuration $\{b\}$ and group elements $g,h$, the commutativity obstruction is only supported on one edge, namely the edge connecting the two vertices $v$ and $2$. When expressed in this form, we can directly see that the commutativity obstruction phase is 1 in the flux-free subspace, as anticipated. The goal of our stabilizer construction is to modify the vertex operators in such a way to remove $\Omega$ and $\Pi$ simultaneously.

\section{Constructing stabilizers from input group $\zz_N$ and a type-I cocycle}\label{app:ZN}
In this section, we will explicitly calculate the obstructions defined above for $G=\zz_N$. We represent the group by the set $\{0,1,\dots,N-1\}$ together with the group multiplication being addition modulo $N$, i.e. $g_1\cdot g_2 =(g_1 + g_2) \mod{N} \eqqcolon g_1\oplus g_2 $. We first introduce operators that generalize the Pauli $Z$ and $X$ matrices to $N$ level systems. These are defined y their action on $\mathcal{H}_l=\spn_\mathbb{C}\{\ket{0},\ket{1},\dots,\ket{N-1}\}$,
\begin{align}
    Z =& \sum_{n=0}^{N-1} \lambda^n \ketbra{n}{n} \qq{with} \lambda\coloneqq e^{2\pi i/N} \quad\qq{and}\quad
    X = \sum_{n=0}^{N-1} \ketbra{n\oplus 1}{n}.
\end{align}
Note that $X^N=Z^N=\mathds{1}$ and that they satisfy the commutation relation
\begin{align}
  XZ = \lambda^{-1} ZX.
\end{align}
Having defined those operators, we can write $A_v^g$ of any $\zz_N$ TQD model in terms of a phase factor (defined in Eq.\,\eqref{eq:vertex_action} by the chosen cocycle $\omega$) times the product $\prod_{l\sim v}X_l^{s(v,l)g}$, where $s(v,l_j)=(-)1$ for an edge $l_j$ pointing (away from)towards vertex $v$. From that, it directly follows that the \textit{phase operator} $Z$ fulfills the following commutativity relation with $A_v^g$:
\begin{align}
    A_v^{g}Z_j = \lambda^{-s(v,l_j)g} Z_jA_v^g ,
\end{align}
where $j$ labels the edge on which $Z_j$ acts and $s(v,l_j)=1~(-1)$ for an edge $l_j$ pointing towards (away from) vertex $v$ (compare Eqs.\,\eqref{eq:vertex_action_edges}). Moreover, we introduce projectors $P^{(n)}_{l_1,\dots,l_k}\coloneqq \sum_{n_1\oplus\dots\oplus n_k = n} \ketbra{n_1,\dots,n_k}$ projecting onto the subspace on which values\footnote{To be precise, we say an edge has a value $l\in\zz_N$ when it is in a state vector $\ket{l}$. Hence, we can sum up the values according to the group multiplication on $\zz_N$.} of the $k$ edges $l_1,\dots,l_k$ sum up to $n\mod N$. Since $n$ is understood modulo $N$, we define $P^{(n+N)}_j =P^{(n)}_j$ for any (set of) edge(s) $j$.

To evaluate the obstruction phases $\Omega$ and $\Pi$ that we have defined in App.\,\ref{app:obstructions}, we first need to choose a (non-trivial) cocycle representative  $[\omega]\in H^3(\zz_N, U(1))$ to insert into Eqs.\,\eqref{eq:app_onsiteobstruction} and \eqref{eq:app_commutativityobstruction}. For $\zz_N$, there are $N$ cocycle classes that can be labelled by $p\in\zz_N$ are represented by
\begin{align}\label{eq:app_ZNcocycle}
    \omega_p(a,b,c) = e^{\frac{2\pi i}{N^2}p a(b+c- [b\oplus c])} = \begin{cases}\lambda^{ap} & b+c\geq N\\
    1 & \text{else}\end{cases}
    \qcomma a,b,c\in\zz_N.
\end{align}
 Such cocycles, depending only on elements from the same group $\zz_N$, are called \textit{type-I} cocycles \cite{propitius1995topological}. All representatives are generated by $\omega_1$ and $(\omega_1)^0=(\omega_1)^N \equiv1$ represents the trivial cocycle class. Investigating the TQD model defined by the cocycle $\omega_1$ and lifting the obstructions for that particular model with a \textit{generating modification phase} is therefore enough to lift the obstructions for all $\zz_N$ TQD models. The modification phase that lifts the obstruction for a model defined with an $\omega_p$ cocycle is given as $p$th power of the generating modification phase.

Inserting $\omega_1$ from Eq.\,\eqref{eq:app_ZNcocycle} into Eq.\,\eqref{eq:app_onsiteobstruction} gives the on-site obstruction
\begin{align}\label{eq:app_ZNonsitephase}
    \Omega_{\{b\}}^{(l_2,l_5)}(g,h) = \left(\lambda^{-B_{3,9,8,1}} \right)^{\Delta_{g+h,N}} \lambda^{g\sum_{i = g}^{g+h-1} \left(P_{4,11,-5}^{(i-l_5)} - P_{6,12,-5}^{(i-l_5)}\right)} = \Omega_{\{b\}}^{(l_5)}(g,h),
\end{align}
where $B_{3,9,8,1}= (-l_3)\oplus(-l_9)\oplus l_8 \oplus l_1$ and $\Delta_{g+h,N} = 1$ for $g+h\geq N$ and $0$ otherwise. Interestingly, the $l_2$ dependence drops out such that, for a fixed flux configuration, the modification phase is only supported on $l_5$. Similarly, inserting $\omega_1$ into Eq.\,\eqref{eq:app_commutativityobstruction} gives the commutativity obstruction
\begin{align}\label{eq:app_ZNcommutativityphase}
    \Pi_{\{b\}}^{(l_2)}(g,h) = \lambda^{g\sum_{i=g-h}^{g-1} (P_{-2,8,1}^{(i-l_2)} - P_{-2,9,3}^{(i-l_2)})}.
\end{align}
Note that these obstructions coincide with the ones calculated for $N=2$, where $g=h=1$ is the only non-trivial case, in Sec.\,\ref{sec:Z2}.

Having quantified the obstructions for the $\zz_N$ TQD model, we introduce the \textit{modified vertex operators} $\Tilde{A}_v^g = D_v^gA_v^g$ with the phase modification $D_v^g$ being a (in the edge basis) diagonal operator with all entries taking values being roots of unity. We impose that $\eval{D_v^g}_{\text{flux-free}} = 1$ so that the ground space properties -- and with them the topological data of the model -- remain unchanged. Let us first consider two vertex operators acting on the same vertex. The modified operators should fulfill
\begin{align}\label{eq:app_grouppropertyvertexoperators}
    \Tilde{A}_v^g\Tilde{A}_v^h = \Tilde{A}_v^{g\oplus h}\qcomma\forall g,h.
\end{align}
Inserting the definition of $\Tilde{A}_v^g$ and Eq.\,\eqref{eq:app_onsiteobstruction} yields the \textit{on-site consistency condition} on the phases $\{\eta_v^g\}$,
\begin{align}\label{eq:app_groupmultiplicationconsistency}
    \eta_v^{g\oplus h} = A_v^g \eta_v^h (A_v^g)^{-1} \eta_v^g \Omega_{\{b\}}^{(l_5)}(g,h).
\end{align}
Since $\eta_v^g$ is diagonal in the edge basis,  conjugation with $A_v^g$ modifies it only by shuffling edge values. In particular, $A_v^g l_2 (A_v^g)^{-1} = l_2\oplus g$ and $A_v^g l_5 (A_v^g)^{-1}= l_5 \oplus(-g)$. Using the cyclic property of $\zz_N$, we can set $g=1$ and $h=N-1$ in Eq.\,\eqref{eq:app_groupmultiplicationconsistency} to obtain an equation for $\eta^0$. Since $A_v^0=\mathds{1}$, $\eta_v^0 = 1$, and we obtain
\begin{align}
    1 = \eta_v^0 = \eta_v^{(N-1)\oplus 1} \stackrel{\eqref{eq:app_groupmultiplicationconsistency}}{=} A_v^1 \eta_v^{N-1} (A_v^1)^{-1} \eta_v^1 \Omega_{\{b\}}^{(l_5)}(1, N-1).
\end{align}
We expand this further by recursively writing $N-1 = (N-2)\oplus 1$, $N-2 = (N-3)\oplus 1,\dots,2=1\oplus1$ and using Eq.\,\eqref{eq:app_groupmultiplicationconsistency} to rewrite $\eta_v^{N-1},\eta_v^{N-2},\dots, \eta_v^{2}$, thereby obtaining the \textit{closure relation}
\begin{align}
\begin{split}
    1 =& \prod_{n=0}^{N-1}A_v^n\Omega_{\{b\}}^{(l_5)}(1,-(n+1))(A_v^n)^{-1} \prod_{n=0}^{N-1} A_v^n \eta_v^1 (A_v^n)^{-1}\\
    =& \prod_{n=0}^{N-1} \Omega_{\{b\}}^{(l_5-n)}(1,-(n+1))  \prod_{n=0}^{N-1} A_v^n \eta_v^1 (A_v^n)^{-1},
\end{split}\label{eq:app_closure}
\end{align}
where we have used that a diagonal operator (such as $\Omega$ and $\eta$) conjugated by $A_v^g$ is still a diagonal operator and therefore commutes with any other diagonal operator. Eq.\,\eqref{eq:app_closure} makes is possible to find a solution for the phase corresponding to the generator of $\zz_N$, $\eta_v^1$. Moreover, the recursion process that led us to Eq.\,\eqref{eq:app_closure} can be used iteratively to generate all other phases $\eta_v^g\;\forall g\in\zz_N$ so that Eq.\,\eqref{eq:app_grouppropertyvertexoperators} is fulfilled. Once that is achieved, the general commutativity problem reduces to restoring  commutativity to the generator $\{\Tilde{A}_v^1\}$ since any other modified vertex operator decomposes as $\Tilde{A}_v^g = (\Tilde{A}_v^1)^g$. We will therefore first solve the on-site consistency condition and then derive a second consistency condition for the commutativity of the modified vertex operators.

Inserting the explicit form of the obstruction phase $\Omega_{\{b\}}^{(l_5)}$ calculated before, the equation reads
\begin{align}
    1 = \lambda^{B_{3,9,8,1}}\lambda^{\sum_{n=0}^{N-1}\sum_{i=1}^{N-1-n} \left( P^{(i-l_5+n)}_{4,11,-5} - P_{6,12,-5}^{(i-l_5+n)}\right)} \prod_{n=0}^{N-1} A_v^n \eta_v^1 (A_v^n)^{-1}.
\end{align}
The double sum in the exponent can be simplified with some projector algebra. It reads
\begin{align}
\begin{split}
    \sum_{n=0}^{N-1}\sum_{i=1}^{N-1-n} \left( P^{(i-l_5+n)}_{4,11,-5} - P_{6,12,-5}^{(i-l_5+n)}\right) =& \sum_{n=0}^{N-1}\sum_{i=1+n}^{N-1} \left( P^{(i-l_5)}_{4,11,-5} - P_{6,12,-5}^{(i-l_5)}\right)\\
    =& -\sum_{n=1}^{N-1}n\left( P^{(n-l_5)}_{4,11,-5} - P_{6,12,-5}^{(n-l_5)} \right) = -B_{6,12,11,4},
\end{split}\label{eq:app_doublesumprojectors}
\end{align}
where we have used that $\sum_{n=0}^{N-1} P^{(n)}_j = \mathds{1}$ and
noting that the final expression is the operator measuring the flux through the
diamond to the right of the vertex, $B_{6,12,11,4} = l_6\oplus l_{12}\oplus(-l_{11})\oplus(-l_4)$. With that, the closure condition
on $\eta_v^1$ reads
\begin{align}\label{eq:app_ZNclosure}
    1 = \lambda^{-B_{3,9,8,1} - B_{6,12,11,4}} \prod_{n=0}^{N-1} A_v^n \eta_v^1 (A_v^n)^{-1}.
\end{align}
Since the vertex operators do not change fluxes, this equation can be solved by any expression of the form
\begin{align}\label{eq:app_ZNonsitesolution}
    \eta_v^1 = \lambda^{(B_{3,9,8,1} + B_{6,12,11,4})/N}\bar{\eta}_v^1\qq{with} \prod_{n=0}^{N-1} A_v^n \bar{\eta}_v^1 (A_v^n)^{-1} = 1.
\end{align}
We are left with a freedom $\bar{\eta}_v^1$ to solve an additional consistency equation coming from the commutativity obstruction phase \eqref{eq:app_ZNcommutativityphase}. To be precise, imposing that the \textit{generating vertex operators} $\{A_v^1\}$ commute, i.e., $\comm*{\Tilde{A}_2^1}{\Tilde{A}_v^1}_G=1$, yields the \textit{commutativity consistency condition} on $\{\eta_v^1\}$
\begin{align}
\begin{split}
    \Pi_{\{b\}}^{(l_2)}(1,1) =& \eta_v^h A_2^1 (\eta_v^h)^{-1}(A_2^1)^{-1}
    A_v^1 \eta_2^1 (A_v^1)^{-1} (\eta_2^1)^{-1}\\
    =& \comm{\eta_v^h}{A_2^1}_G\comm{A_v^1}{\eta_2^1}_G.
\end{split}
\end{align}
Substituting in Eqs.\,\eqref{eq:app_ZNonsitesolution} and \eqref{eq:app_ZNcommutativityphase}, we obtain the two consistency conditions
\begin{align}
    \lambda^{P_{-2,8,1}^{(-l_2)} - P_{-2,9,3}^{(-l_2)}} = \bar{\eta}_v^1 A_2^1 (\bar{\eta}_v^1)^{-1}(A_2^1)^{-1}
 A_v^1\bar{\eta}_2^1(A_v^1)^{-1} (\bar{\eta}_2^1)^{-1}\qq{and} \prod_{n=0}^{N-1} A_v^n \bar{\eta}_v^1 (A_v^n)^{-1} = 1
\end{align}
to lift the on-site and commutativity obstructions. We find a one-parameter family of solutions
\begin{align}
    \bar{\eta}_v^1(p) = \lambda^{-(p B_{3,9,8,1} + (1-p)B_{6,12,11,4})/N}\lambda^{-p \sum_{n=0}^{-l_2-1} \left( P_{-2,8,1}^{(n)} - P_{-2,9,3}^{(n)} \right) - (1-p)\sum_{m=0}^{-l_5} \left( P_{6,12,-5}^{(m)} - P_{4,11,-5}^{(m)} \right)},
\end{align}
where we have used a similar manipulation as in Eq.\,\eqref{eq:app_doublesumprojectors} to show the second condition for $\bar{\eta}_v^1$. Note that $\eta_v^1(p+N^2) = \eta_v^1(p)$, and thus all distinct solutions in this family are labeled by $p\in[0,N^2)$. Putting these together with Eq.\,\eqref{eq:app_ZNonsitesolution} yields the full modification phase
\begin{align}
    \eta_v^1(p) = \lambda^{((1-p) B_{3,9,8,1} + pB_{6,12,11,4})/N}\lambda^{-p \sum_{n=0}^{-l_2-1} \left( P_{-2,8,1}^{(n)} - P_{-2,9,3}^{(n)} \right) - (1-p)\sum_{m=0}^{-l_5} \left( P_{6,12,-5}^{(m)} - P_{4,11,-5}^{(m)} \right)}.
\end{align}
This expression consists of two parts. The rightmost one, built from two sums of projectors and having an explicit $l_2$ and $l_5$ dependence ensures commutativity, whereas the other part, depending only on fluxes and therefore not altering the commutativity properties, is there to fulfill the on-site condition. Using the recurrence relation from the derivation of the closure relation Eq.\,\eqref{eq:app_groupmultiplicationconsistency}, we generate all other modification phases $\{\eta_v^g\}$ iteratively so that the on-site condition is fulfilled. The general modification phase then reads
\begin{align}\label{eq:app_ZN_solutionfamily}
\begin{split}
    \eta_v^g(p) =& \lambda^{g((1-p) B_{3,9,8,1} + pB_{6,12,11,4})/N} \lambda^{-g\left[p\sum_{n=0}^{-l_2-g}\left( P_{-2,8,1}^{(n)} - P_{-2,9,3}^{(n)}\right) + (1-p)\sum_{n=0}^{-l_5} \left( P_{6,12,-5}^{(n)} - P_{4,11,-5}^{(n)} \right)\right]}\\
    &\times \lambda^{-\sum_{n=0}^{g-1}n\left[ p \left( P_{-2,8,1}^{(-l_2-n)} - P_{-2,9,3}^{(-l_2-n)}\right) +  (2-p) \left( P_{6,12,-5}^{(-l_5+n)} - P_{4,11,-5}^{(-l_5+n)} \right)\right]}\qcomma p\in\mathbb{R}.
\end{split}
\end{align}
With that, we have obtained a solution for a very general case, namely all $\zz_N$ TQD models. On a first glance this expression seems complex, but a little inspections shows that, for specific choices of $p$ and certain (small) local dimensions, it reduces to a manageable expression (see Sec.\,\ref{sec:Z2}).

\section{Constructing stabilizers from input group $\zz_N\times \zz_N$ and a type-II cocycle}\label{app:ZNZN}
In this section, we will calculate the obstruction phases derived in App.\,\ref{app:obstructions} for $G=\zz_N^2$ and a \textit{type-II} cocycle. Each element in the input group can be written as a pair of $\zz_N$ elements, $g=(g_1,g_2)$ with group multiplication naturally carrying over from $\zz_N$.\footnote{We use the same symbol ``$\oplus$" for the group multiplication on $\zz_N$ and $\zz_N^2$.} It is clear that this group is generated by two elements, namely $(0,1)$ and $(1,0)$. When looking at the cohomology of this group, one finds that the resulting cocycle classes are generated by three elements split into two \textit{types}\cite{propitius1995topological}. The two \textit{type-I} generators are the same as those for $\zz_N$, depending on the data from a single tensor factor, i.e.,
\begin{align}
    \omega_{I,p}(a,b,c) = \omega_p(a_i,b_i,c_i)\qcomma i=1,2,
\end{align}
where $\omega_p$ was defined in Eq.\,\eqref{eq:app_ZNcocycle}. Using such a cocycle to define a TQD model will result in the same functional form of the obstructions $\Omega$ and $\Pi$ found for $\zz_N$. Hence, they also can be removed by the same modification phase $\eta_v^g$ from Eq.\,\eqref{eq:app_ZN_solutionfamily} where all $\zz_N$ variables now carry a tensor factor index $i$.


Besides these type-I cocycles, there are \textit{type-II} cocyles that depend on both tensor factors simultaneously and can be represented by
\begin{align}\label{eq:app_ZNZNtype2cocycle}
    \omega_{II,p}(a,b,c) = \omega_p(a_1,b_2,c_2) = \begin{cases}\lambda^{a_1 p} & b_2+c_2\geq N\\
    1 & \text{else}\end{cases},
\end{align}
where $\lambda=e^{2\pi i/N}$. One could also define the cocycle with indices 1 and 2 interchanged, but this is known to be gauge equivalent to the above definition \cite{propitius1995topological}.
Eq.\,\eqref{eq:app_ZNZNtype2cocycle} shows that cocycles of type II mixes the two tensor factors of the input group elements in a non-trivial way. Whereas the $\zz_N^2$ TQD model with a type-I cocycle can be decomposed into two (possibly inequivalent) $\zz_N$ TQD models, a type-II cocyle gives rise to a different topological order that cannot be factored in this way. In the following, we calculate the obstructions $\Omega$ and $\Pi$ with a type-II cocycle and investigate how to lift these with appropriately chosen phase modifications.

Inserting the chosen type-II cocycle representative $\omega_{II,1}$ from Eq.\,\eqref{eq:app_ZNZNtype2cocycle} into the obstruction phases calculated in Eqs.\,\eqref{eq:app_onsiteobstruction} and \eqref{eq:app_commutativityobstruction}, we find that the vertex operators of the type-II $\zz_N^2$ TQD model fail to implement the group action faithfully on site, generating the obstruction phase
\begin{align}
\begin{split}
    \Omega_{\{b\}}^{(l_5)}(g,h) =& \lambda^{-B_{3,9,8,1}^{(2)}\Delta_{g^{(1)}+h^{(1)},N} - B_{3,9,8,1}^{(1)}\Delta_{g^{(2)}+h^{(2)},N}}\\
    &\times \lambda^{g^{(2)}\sum_{i = g^{(1)}}^{g^{(1)}+h^{(1)}-1} \big(P_{4,11,-5^{(1)}}^{(i- l_5^{(1)})} - P_{6,12,-5^{(1)}}^{(i- l_5^{(1)})}\big)}%
    \lambda^{g^{(1)}\sum_{i = g^{(2)}}^{g^{(2)}+h^{(2)}-1} \big(P_{4,11,-5^{(2)}}^{(i- l_5^{(2)})} - P_{6,12,-5^{(2)}}^{(i- l_5^{(2)})}\big)},
\end{split}
\end{align}
where the fluxes and projectors are defined as in the previous section but with every group element and edge value carrying an additional (upper) index $(i)$ labelling the tensor factor of the corresponding variable. To avoid notation clutter in the projectors, we only write the tensor factor index once.  For example, $P_{4,11,-5^{(1)}}^{(n)}$ projects onto the subspace where $l_4^{(1)}\oplus l_{-1}^{(1)}\oplus(-l_5^{(1)}) = n$. The quantity $\Delta_{\bullet,\bullet}$ is defined as in Eq.\,\eqref{eq:app_ZNonsitephase}. Note that this obstruction phase consists of similar terms as those found for $\zz_N$ case, but with added mixing between tensor factors. The analoguous calculation of the commutativity obstruction phase yields
\begin{align}\label{eq:app_ZNZNcommutativityobstruction}
    \Pi_{\{b\}}^{(l_2)}(g,h) = \lambda^{g^{(2)} \sum_{i=g^{(1)}-h^{(1)}}^{g^{(1)}-1} \big(P_{-2,8,1^{(1)}}^{(i-l^{(1)}_2)} - P_{-2,9,3^{(1)}}^{(i-l^{(1)}_2)}\big)}\lambda^{g^{(1)} \sum_{i=g^{(2)}-h^{(2)}}^{g^{(2)}-1} \big(P_{-2,8,1^{(2)}}^{(i-l^{(2)}_2)} - P_{-2,9,3^{(2)}}^{(i-l^{(2)}_2)}\big)}.
\end{align}

The procedure to lift these obstructions begins identically to that in the previous section. Again, we derive \textit{closure relations} from group multiplication in every cyclic sub-factor of $\zz_N^2$. In addition to these, one also finds extra constraint equations coming from group multiplication of elements from different sub-factors when the model includes a non-trivial type-II cocycle. Just as in the previous section, once the group multiplication is implemented consistently on-site, we only need to make the vertex operators for the generators $(0,1)$ and $(1,0)$ commute and can then iteratively construct all other modification phases so that both the on-site and commutativity obstructions are removed.

We start by introducing the modified vertex operators $\Tilde{A}_v^g = \eta_v^gA_v^g\;\forall g\in\zz_N^2$ and imposing Eq.\,\eqref{eq:app_grouppropertyvertexoperators}. In particular, this directly implies a general condition on the modification phases, Eq.\,\eqref{eq:app_groupmultiplicationconsistency}. This enables us to use cyclicity to iteratively derive the three consistency conditions from the relations $(N-1,0)\oplus(1,0) = (0,0)$, $(0,N-1)\oplus(0,1) = (0,0)$ and $(N-1,N-1)\oplus(1,1) = (0,0)$, yielding
\begin{subequations}\label{eq:app_ZNZNcolsurerelations}
\begin{align*}
    1 =& \eta_v^{(0,0)} = \eta_v^{(1,0)\oplus\dots\oplus(1,0)} = \prod_{n=0}^{N-1}A_v^{(n,0)}\Omega_{\{b\}}^{(l_5)}((1,0),(-(n+1),0))(A_v^{(n,0)})^{-1}\prod_{m=0}^{N-1} A_v^{(m,0)}\eta_v^{(1,0)}(A_v^{(m,0)})^{-1}\\
    =& \lambda^{-B_{\text{3,9,8,1}}^{(2)}}\prod_{m=0}^{N-1} A_v^{(m,0)}\eta_v^{(1,0)} (A_v^{(m,0)})^{-1},\numberthis\label{eq:app_ZNZNconsistencyA}\\
    1 =& \eta_v^{(0,0)} = \eta_v^{(0,1)\oplus\dots\oplus(0,1)} = \prod_{n=0}^{N-1} A_v^{(0,n)}\Omega_{\{b\}}^{(l_5)}((0,1),(0,-(n+1)))(A_v^{(0,n)})^{-1}\prod_{m=0}^{N-1}A_v^{(0,m)}\eta_v^{(0,1)}(A_v^{(0,m)})^{-1}\\
    =& \lambda^{-B_{3,9,8,1}^{(1)}} \prod_{m=0}^{N-1}A_v^{(0,m)}\eta_v^{(0,1)}(A_v^{(0,m)})^{-1}\qq{and}\numberthis\label{eq:app_ZNZNconsistencyB}\\
    1 =& \eta_v^{(0,0)} = \eta_v^{(1,1)\oplus\dots\oplus(1,1)} = \prod_{n=0}^{N-1}A_v^{(n,n)}\Omega_{\{b\}}^{(l_5)}((1,1),(-(n+1),-(n+1)))(A_v^{(n,n)})^{-1}\prod_{m=0}^{N-1}A_v^{(n,n)}\eta_v^{(1,1)}(A_v^{(n,n)})^{-1}\\
    \stackrel{\eqref{eq:app_groupmultiplicationconsistency}}{=}& \prod_{n=0}^{N-1}A_v^{(n,n)}\Omega_{\{b\}}^{(l_5)}((1,1),(-(n+1),-(n+1))(A_v^{(n,n)})^{-1}\\
    &\times\prod_{m=0}^{N-1}A_v^{(m,m)}A_v^{(1,0)}\eta_v^{(0,1)}(A_v^{(1,0)})^{-1}\eta_v^{(1,0)}\Omega((1,0),(0,1)) (A_v^{(m,m)})^{-1}\\
    =&\lambda^{-\sum_{i=1}^2\left( B_{3,9,8,1}^{(i)} + B_{6,12,11,4}^{(i)}\right)}\prod_{m=0}^{N-1}A_v^{(m+1,m)}\eta_v^{(0,1)}(A_v^{(m+1,m)})^{-1} A_v^{(m,m)}\eta_v^{(1,0)}(A_v^{(m,m)})^{-1}.\numberthis\label{eq:app_ZNZNconsistencyC}
\end{align*}
\end{subequations}
The final pre-factors from Eqs.\,\eqref{eq:app_ZNZNconsistencyA}, \eqref{eq:app_ZNZNconsistencyB} and \eqref{eq:app_ZNZNconsistencyC}, where the root of unity $\lambda$ appears with fluxes in the exponent, are produced by the obstruction phases using the same identities as in Eq.\,\eqref{eq:app_doublesumprojectors}.
The first two conditions are analogous to the one found in Eq.\,\eqref{eq:app_ZNclosure}, and so we begin by solving the first two equations Eq.\,\eqref{eq:app_ZNZNconsistencyA} and \eqref{eq:app_ZNZNconsistencyB} in a similar fashion as the closure relation in the $\zz_N$ case, Eq.\,\eqref{eq:app_ZNclosure}. We find that they are easily solved by
\begin{subequations}
\begin{align}
    \eta_v^{(1,0)} =& \lambda^{B_{3,9,8,1}^{(2)}/N}\bar{\eta}_v^{(1,0)}\qq{and}\\
    \eta_v^{(0,1)} =& \lambda^{B_{3,9,8,1}^{(1)}/N}\bar{\eta}_v^{(0,1)},
\end{align}
\end{subequations}
where $\prod_{m=0}^{N-1}A_v^{(m,0)}\bar{\eta}^{(1,0)}(A_v^{(m,0)})^{-1} = \prod_{m=0}^{N-1}A_v^{(0,m)}\bar{\eta}^{(0,1)}(A_v^{(0,m)})^{-1} = 1$. Note that these two constraints on $\bar{\eta}^{(1,0)}$ and $\bar{\eta}^{(1,0)}$ are fulfilled by any term that is a $N$th root of unity and only depends on the second respectively the first tensor factor of the link variables. Inserting this ansatz into the third closure relation Eq.\,\eqref{eq:app_ZNZNconsistencyC}, we obtain a closure relation for $\bar{\eta}$,
\begin{align*}
  1 = \lambda^{- B_{6,12,11,4}^{(1)} - B_{6,12,11,4}^{(2)}}\prod_{m=0}^{N-1}A_v^{(m+1,m)}\bar{\eta}_v^{(0,1)}(A_v^{(m+1,m)})^{-1}A_v^{(m,m)}\bar{\eta}_v^{(1,0)}(A_v^{(m,m)})^{-1}.
\end{align*}
Given the large freedom available when fulfilling the first two closure relations, we can construct $\bar{\eta}^{(1,0)}$ such that it cancels out $\lambda^{-B_{6,12,11,4}^{(2)}}$ and $\bar{\eta}^{(0,1)}$ such that it cancels out $\lambda^{-B_{6,12,11,4}^{(1)}}$. One solution is given by
\begin{subequations}
\begin{align}
  \bar{\eta}^{(1,0)} =& \lambda^{-\sum_{i=0}^{-l_5^{(2)}-1}\big(P_{6,12,-5^{(2)}}^{(i)}-P_{4,11,-5^{(2)}}^{(i)}\big)}\qq{and}\\
  \bar{\eta}^{(0,1)} =& \lambda^{-\sum_{i=0}^{-l_5^{(1)}-1}\big(P_{6,12,-5^{(2)}}^{(i)}-P_{4,11,-5^{(2)}}^{(i)}\big)}
\end{align}
\end{subequations}
since the tensor factors of $l_5$ are shifted by $m$ and $m+1$, respectively, by the conjugation with $A_v^{(m,m)}$ and $A_v^{(m+1,m)}$, respectively. Then, summing over all $m$ in the exponent (due to the product over all $m$) as in Eq.\,\eqref{eq:app_doublesumprojectors} gives exactly the desired flux.

The full modification phases for the generators therefore read
\begin{subequations}\label{eq:app_ZNZNgeneratingcorrectionphases}
\begin{align}
  \eta^{(1,0)} =& \lambda^{B_{3,9,8,1}^{(2)}/N}\lambda^{-\sum_{i=0}^{-l_5^{(2)}-1}\big(P_{6,12,-5^{(2)}}^{(i)}-P_{4,11,-5^{(2)}}^{(i)}\big)}\label{eq:app_ZNZN_generatingphasesa},\\
  \eta^{(0,1)} =& \lambda^{B_{3,9,8,1}^{(1)}/N}\lambda^{-\sum_{i=0}^{-l_5^{(1)}-1}\big(P_{6,12,-5^{(2)}}^{(i)}-P_{4,11,-5^{(2)}}^{(i)}\big)}\label{eq:app_ZNZN_generatingphasesb}.
\end{align}
\end{subequations}
Interestingly, the phase for the first generator only depends on the second tensor factor of fluxes and edges and vice versa. This is a direct consequence of the way the type-II cocycle couples the two tensor factors.

The other modification phases can be calculated iteratively using Eq.\,\eqref{eq:app_groupmultiplicationconsistency} to produce a proper representation of the group action. We will give an explicit expression the modification phases for any group element $g\in\zz_N^2$ at the end of this section after having discussed the commutativity obstruction.
Once the vertex operators faithfully represent the group action on-site, every modified vertex operator can be decomposed in terms of the generating vertex operators $\Tilde{A}_v^{(0,1)}$ and $\Tilde{A}_v^{(1,0)}$. Therefore, it is sufficient to resolve the commutativity obstruction for those two operators while still fulfilling Eqs.\,\eqref{eq:app_ZNZNcolsurerelations}.
Imposing $\comm*{\Tilde{A}_2^g}{\Tilde{A}_v^h}_G = \mathds{1}\;\forall\,g,h\in\{(0,1);(1,0)\}$ and evaluating the obstruction phases $\Pi_{\{b\}}^{(l_2)}(g,h)$ for the corresponding $g,h$ yields the conditions
\begin{subequations}
\begin{align}
    A_2^{(1,0)}(\eta_v^{(1,0)})^{-1}(A_2^{(1,0)})^{-1}\eta_v^{(1,0)} A_v^{(0,1)}\eta_2^{(1,0)}(A_v^{(0,1)})^{-1}(\eta_2^{(1,0)})^{-1} =& \Pi_{\{b\}}^{(l_2)}((1,0),(1,0)) \stackrel{\eqref{eq:app_ZNZNcommutativityobstruction}}{=} 1 ,\\
    A_2^{(0,1)}(\eta_v^{(0,1)})^{-1}(A_v^{(0,1)})^{-1}\eta_v^{(0,1)} A_v^{(0,1)}\eta_2^{(0,1)}(A_v^{(0,1)})^{-1}(\eta_2^{(0,1)})^{-1} =& \Pi_{\{b\}}^{(l_2)}((0,1),(0,1)) \stackrel{\eqref{eq:app_ZNZNcommutativityobstruction}}{=} 1,\\
    A_2^{(1,0)}(\eta_v^{(0,1)})^{-1}(A_2^{(1,0)})^{-1}\eta_v^{(0,1)} A_v^{(0,1)}\eta_2^{(1,0)}(A_v^{(0,1)})^{-1}(\eta_2^{(1,0)})^{-1} =& \Pi_{\{b\}}^{(l_2)}((1,0),(0,1)) \stackrel{\eqref{eq:app_ZNZNcommutativityobstruction}}{=} \lambda^{P_{-2,8,1^{(2)}}^{(-l_2^{(2)}-1)} - P_{-2,9,3^{(2)}}^{(-l_2^{(2)}-1)} },\\
     A_2^{(0,1)}(\eta_v^{(1,0)})^{-1}(A_2^{(0,1)})^{-1}\eta_v^{(1,0)} A_v^{(1,0)}\eta_2^{(0,1)}(A_v^{(1,0)})^{-1}(\eta_2^{(0,1)})^{-1} =& \Pi_{\{b\}}^{(l_2)}((0,1),(1,0)) \stackrel{\eqref{eq:app_ZNZNcommutativityobstruction}}{=} \lambda^{P_{-2,8,1^{(1)}}^{(-l_2^{(1)}-1)} - P_{-2,9,3^{(1)}}^{(-l_2^{(1)}-1)} }.
\end{align}
\end{subequations}
Surprisingly, the generating modification phases derived from the on-site condition (see Eqs.\,\eqref{eq:app_ZNZNgeneratingcorrectionphases}) also fulfill these four equations. As mentioned, Eq.\,\eqref{eq:app_groupmultiplicationconsistency} now allows us to iteratively generate all modification phases for any group element $g=(g_1,g_2)$. It reads
\begin{align}
  \eta_v^g = \lambda^{(g_1 B_{3,9,8,1}^{(2)}+ g_2 B_{3,9,8,1}^{(1)})/N} \lambda^{-g_1 \sum_{i=0}^{-l_5^{(2)}-1+g_2}\big(P_{6,12,-5^{(2)}}^{(i)}-P_{4,11,-5^{(2)}}^{(i)}\big) -g_2 \sum_{i=0}^{-l_5^{(1)}-1+g_1}\big(P_{6,12,-5^{(1)}}^{(i)}-P_{4,11,-5^{(1)}}^{(i)}\big)}.\label{eq:app_ZNZNcorrectionphase}
\end{align}
With that, we have resolved both obstructions for the $\zz_N^2$ TQD model with a type-II cocycle.
\end{document}